# Interplay between Interlayer Shift and Twist: Twisted van der Waals Nanowires Driven by Rotational Twinning

*Dong-gyu Kim[1,†], Kisung Kang[2,3,†], Hani Kang[1,†], Kihyun Lee[1], Yangjin Lee[1,4], Jinsub Park[1], Joong-Eon Jung[1], Min Kim[1], Myeongjin Jang[1], Aloysius Soon[2,*], and Kwanpyo Kim[1,*]*

[1]Department of Physics, Yonsei University, Seoul 03722, Republic of Korea.

[2]Department of Materials Science & Engineering, Yonsei University, Seoul 03722, Republic of Korea

[3]School of Materials Science and Engineering, Chonnam National University, Gwangju 61186, Republic of Korea

[4]Institute of Advanced Composite Materials, Korea Institute of Science and Technology (KIST), Jeonbuk 55324, Korea

*Corresponding Authors: A.S. (aloysius.soon@yonsei.ac.kr), K.K. (kpkim@yonsei.ac.kr)

[†]Equal Contributions

**Abstract**

**In van der Waals (vdW) layered materials, interlayer shift and twist have enabled the control of material properties through polytype and moiré engineering. Various approaches, including bottom-up synthesis and manual layer-by-layer stacking, have been utilized to engineer targeted stacking configurations. However, the interplay between interlayer shift and twist, as well as reliable mechanisms for fine-tuning these parameters, remains largely unexplored. Here, we report a previously unrecognized twisting mechanism arising from preferred tilted stacking and twinning in vdW crystals. Electron diffraction and atomic-resolution scanning transmission electron microscopy (STEM) imaging reveal that the lattice planes of group-IV chalcogenide $GeSe_{2-x}Te_x$ rotate continuously along the nanowire growth axis, with twist rates depending systematically on nanowire radius. Atomic-scale imaging further identifies a continuous rotational twin boundary extending along the central region of the nanowire. First-principles calculations and structural relaxation simulations confirm that the twisting deformation originates from energetic competition between the preferred interlayer stacking registry and the strain cost imposed by rotational twinning. These findings establish rotational twinning as an intrinsic route to spontaneous twist formation and provide a design principle for realizing twist-engineered vdW crystals with compatible crystal symmetries and stacking motifs.**

## Introduction

Twisting provides a route to tuning electrical, optical, and mechanical responses in condensed matter systems. Crystals that twist during growth have been observed across a broad range of organic and inorganic materials [1]. In organic molecular crystals and semicrystalline polymers, surface stresses have been proposed as a driving force for twisting [2,3], while impurity-induced lattice strain has also been implicated in molecular crystals [4]. In inorganic crystals, proposed mechanisms include twinning, dislocations, and inhomogeneous strain [5-7].

Twisting between adjacent layers in two-dimensional (2D) van der Waals (vdW) materials has emerged as a powerful degree of freedom for tuning material properties and inducing correlation-driven phenomena [8,9]. Twistronics has demonstrated that small interlayer rotations can generate moiré superlattices [10] and induce emergent properties, including correlated insulating states [11], superconductivity [12], and moiré excitons and related excitonic responses [13-15], establishing the twist angle as a structural degree of freedom. Producing twisted vdW crystals has mainly relied on layer-by-layer stacking fabrication or bottom-up synthesis [9,16-18]. In particular, some vdW nanowires have shown continuous twisting along the growth direction, which can be described by the classical Eshelby twist mechanism driven by a screw dislocation at the core [17-20]. However, alternative intrinsic crystallographic mechanisms capable of producing continuous lattice twisting remain largely unexplored.

Interlayer shift provides another important structural degree of freedom in vdW crystals [21,22]. Polytype transitions driven by temperature, electric field, and doping have demonstrated that changes in interlayer registry can substantially modify material properties and enable new functionalities [23-27]. Still, the interplay between interlayer shift and twist, as well as reliable mechanisms for fine-tuning these parameters, remains largely unexplored in vdW crystals. In particular, whether an energetic preference for a specific interlayer stacking registry can couple with a twin structure and drive continuous lattice rotation has not been well established.

Realizing twisted crystals based on stacking registry could therefore provide a design principle for new twist-engineered vdW systems, offering opportunities to explore novel functionalities [28].

Here, we report the interplay between interlayer shift and twist in vdW crystals, uncovering a previously unrecognized twisting mechanism. The nanowires based on group-IV chalcogenide $GeSe_{2-x}Te_x$ display continuous twisting along the nanowire growth axis, with twist rates depending systematically on nanowire radius. First-principles calculations and structural relaxation simulations confirm that the twisting deformation originates from energetic competition between the preferred interlayer stacking registry and the strain cost imposed by rotational twinning. Our results suggest that this mechanism may extend to vdW materials with comparable unit-cell symmetries and stacking motifs, providing a general structural principle for twist-engineered layered systems.

**Results and Discussion**

$GeSe_{2-x}Te_x$ nanowires were synthesized by Au-catalyzed vapor–liquid–solid (VLS) growth (Fig. S1). In VLS growth, precursor incorporation into a liquid catalyst and supersaturation at the catalyst–solid interface govern one-dimensional crystal growth, while catalyst size, wetting, undercooling, and substrate interactions affect the final morphology [29,30]. Previous studies have demonstrated diverse crystal morphologies in vapor-phase growth of layered chalcogenides, including nanoribbons and anisotropic flakes [31,32]. The elemental ratio of Ge and Se and minute incorporation of Te were confirmed by Energy-Dispersive X-ray Spectroscopy (EDS) of nanowires (Fig. S2d).

Raman spectroscopy reveals that the vibrational spectra of the nanowires closely resemble those of β-$GeSe_2$ (Fig. S2). β-$GeSe_2$ crystallizes in a monoclinic $P2_1/c$ space group with reported lattice parameters of a = 7.016 Å, b = 16.796 Å, c = 11.831 Å, and β = 90.65° [33].

Previous studies have reported bandgap values of $GeSe_2$ in the range of approximately 2.5–2.7 eV [34,35], consistent with the cathodoluminescence response observed from the synthesized nanowires (Fig. S3).

Position-resolved selected-area electron diffraction (SAED) patterns acquired along individual nanowires reveal systematic changes in crystal orientation. Although the diffraction peaks associated with the nanowire axis remain fixed at all measured positions, surrounding peaks rotate with position (Fig. 1a), indicating lattice rotation about the growth axis rather than random local deformation. Wire-tracking diffraction, simulated diffraction patterns, and tilting experiments were used to identify low-index zone axes and reconstruct the three-dimensional reciprocal lattice (Fig. S4). The refined unit cell is triclinic, with lattice parameters a = 7.0 Å, b = 17.0 Å, c = 7.55 Å, $\alpha$ = 110°, $\beta$ = 117°, and $\gamma$ = 90° (Fig. 1b, c). Compared with the previously known monoclinic $\beta$-$GeSe_2$ structure, the nanowires adopt a distinct triclinic sequential-stacking structure, as further confirmed by crystallographic comparison (Table S1). Based on this refined unit cell, simulated diffraction-pattern evolution during lattice rotation further supports that the observed systematic spot displacement originates from continuous lattice rotation (Fig. S5). This sequential-stacking arrangement reduces the symmetry relative to bulk $\beta$-$GeSe_2$. Consequently, the refined triclinic nanowire unit cell lacks inversion symmetry.

The structural difference originates from a stacking transformation of $\beta$-$GeSe_2$-like building blocks. Relative in-plane sliding between adjacent $\beta$-$GeSe_2$-like layers converts the AB-type bulk arrangement into a sequential-stacking configuration (Fig. 1d). Sequential stacking refers to a structure in which each vdW layer undergoes a consistent in-plane translation relative to the underlying layer, producing a cumulative lateral shift rather than a fixed AA or AB registry. The corresponding unit-cell contents are also reduced from 48 atoms in the $\beta$-$GeSe_2$ unit cell to 24 atoms in the triclinic nanowire unit cell. Density functional theory (DFT) calculations show that the sequential-stacking configuration is energetically competitive with bulk β-$GeSe_2$, with only a small energy difference (Fig. S6).

The position-dependent diffraction analysis further shows that lattice orientation evolves continuously along the nanowire length. Distance–twist angle measurements reveal gradual lattice rotation, while radius–twist rate analysis demonstrates that nanowires with smaller radii exhibit larger twist rates (Fig. 1e and Fig. S7). Dark-field TEM tilt-series imaging further supports the continuous twisting of the nanowire by visualizing position-dependent contrast changes during tilting in a representative nanowire (Fig. S8). Raman mapping also provides additional evidence for orientation modulation along the nanowire axis, as reflected by periodic changes in the $A_g^1/A_g^2$ intensity ratio (Fig. S9). Together, these results establish that the synthesized nanowires possess a triclinic sequential-stacking structure and undergo continuous twisting with a radius-dependent twist rate.

Atomic-resolution STEM imaging was performed on nanowire regions with pronounced kinked morphology to identify the structural origin of twisting. Surprisingly, images acquired along the [100] zone axis reveal a rotational twin boundary extending through the central region of the nanowire, separating two crystalline domains (Fig. 2a). The rotational twin boundary is not readily resolved by electron diffraction because the diffraction patterns from the two rotationally related domains largely overlap under the measured zone-axis conditions (Fig. S10). Across this boundary, the layers adopt a sequential-stacking configuration, and the triclinic c-axis is tilted by approximately 22° from the nanowire growth axis. Importantly, no screw dislocation is observed in the central region of the nanowire, distinguishing this structure from conventional Eshelby-twist nanowires [17,18]. After identifying the rotational twin boundary as the structural origin of twisting, handedness analysis further showed that right- and left-handed nanowires coexist without a clear radius dependence, suggesting stochastic handedness selection during twin-boundary formation (Fig. S11).

To clarify the stacking configuration, STEM image simulations were performed for both monoclinic β-$GeSe_2$ and the synthesized nanowire structure. In simulated β-$GeSe_2$ images, Ge–Se clusters form a zigzag arrangement characteristic of AB-type stacking, whereas in the nanowire model, the clusters align along a single line, reflecting sequential stacking (Fig. 2b). This contrast confirms that the synthesized nanowires possess a stacking sequence distinct from that of β-$GeSe_2$.

Cross-sectional STEM analysis was then performed through the central twin-boundary region (Fig. 2c). The imaging direction was assigned to the [102] zone axis of the refined triclinic unit cell by comparing the experimental diffraction pattern with the corresponding simulated pattern (Fig. 2d). Atomic-resolution images acquired along this direction show structural motifs consistent with the refined model, as further supported by STEM image simulations (Fig. 2e). A comparison of the Ge–Se cluster motifs shows that the motif in grain 2 matches the rotated configuration of grain 1 rather than its mirror-reflected counterpart, supporting the assignment of a rotational twin boundary (Fig. 2f). These observations establish that the synthesized nanowires contain a continuous rotational twin boundary along the central region of the nanowire.

First-principles DFT calculations were performed to evaluate the energetic stability of this rotational-twin-based stacking structure. A periodic supercell containing two rotational twin boundaries was constructed from the sequential-stacking structure, producing a V-shaped stacking geometry consistent with the STEM observations (Fig. 2g). The twin-boundary formation energy was defined as

$$E_{\mathrm{twin}} = \frac{E_{\mathrm{twinned}} - 2NE_{\mathrm{SS}}}{2A_{\mathrm{twin}}}, \quad (1)$$

where $E_{\mathrm{twinned}}$ is the total energy of the supercell containing two rotational twin boundaries, $E_{\mathrm{SS}}$ is the energy of a sequential-stacking unit cell, $N$ is the number of repeated unit cells, and $A_{\mathrm{twin}}$ is the area of one rotational twin boundary in the periodic supercell. To minimize artificial interactions between periodically repeated twin boundaries, the formation energy was extrapolated with increasing supercell size to estimate the isolated boundary limit. The calculated formation energy remains low over the examined range and converges to approximately 0.00083 eV/$\text{Å}^2$ in the isolated boundary limit (Fig. S12). This low formation energy indicates that the rotational twin boundary is an energetically stable structural feature that can form spontaneously in the nanowires.

To examine whether the experimentally observed twisting can arise from a rotational twin boundary alone, structural relaxation simulations were performed using a machine-learning interatomic potential. The initial model was constructed from a DFT-derived layered structure containing a rotational twin boundary but no initial twisting. A cylindrical nanowire geometry was generated by cutting the structure into stacked inverted-V-shaped disks, consistent with STEM observations, and dangling bonds at the disk edges were passivated with hydrogen atoms (Fig. 3a, left). Structural relaxation of this initially untwisted $GeSe_2$ nanowire shows that, even without a screw dislocation, the vdW layers gradually rotate and develop a continuous layer-by-layer twist (Fig. 3a). This result indicates that rotational twinning and interlayer interactions are sufficient to induce twisting. The simulations also reproduce the radius dependence of twisting: nanowires with radii of 1.6 and 2.0 nm exhibit twist rates of approximately −1850 and −910° $\mu m^{-1}$, respectively (Fig. S13). The negative sign reflects the direction of lattice rotation under the chosen angular convention, whereas the magnitude of the twist rate decreases with increasing radius. We also confirmed that random Te incorporation in $GeSe_2$ does not influence its twisting behavior (Fig. S14).

To understand the origin of this twisting behavior, we calculate the interlayer stacking energy as a function of interlayer rotational angle, $\Phi_{\mathrm{t}}$. Here, we consider a rigid relative rotation between the inverted-V-stacked layers about the center of the wire (Fig. 3b). Before the introduction of interlayer rotation, the two layers are positioned directly on top of each other owing to the constraint imposed by the twin boundary. The interlayer stacking configuration projected onto the wire cross-section is given as $(\delta_x = 0, \delta_y = 0)$. Upon interlayer rotation, the local stacking configuration changes to $(\delta_x = -r\Phi_t \sin\theta, \delta_y = r\Phi_t \cos\theta)$ (Fig. 3c,d), where the local coordinates in the wire cross-section are specified by the radial position $r$ and the azimuthal angle $\theta$. Based on the experimentally observed bulk unit-cell geometry, the ground-state stacking configuration requires a relative shift $(\delta_x = 0, \delta_y = \pm\delta_0)$ (Fig. 3e), with the displacement direction parallel to the twin boundary.

Therefore, the local interlayer stacking energy can be expressed as

$$\varepsilon^{\text{stacking}}(\delta_x, \delta_y) = \alpha(\delta_x)^2 + \beta(\delta_y - \delta_0)^2 \tag{2}$$

where $\alpha = 6.77\ \text{meV}/(\text{f.u.}\cdot\text{Å}^2)$, $\beta = 8.52\ \text{meV}/(\text{f.u.}\cdot\text{Å}^2)$, and $\delta_0 = 0.257$ Å were numerically determined using DFT calculations, as shown in Figure 3e-h. After twisting by an angle $\Phi_t$, the local stacking energy at $r$ and $\theta$ becomes

$$\varepsilon^{\text{stacking}}(r, \theta, \Phi_t) = \alpha(r\Phi_t \sin\theta)^2 + \beta(r\Phi_t \cos\theta - \delta_0)^2 \tag{3}$$

The total stacking energy is obtained by integrating the local energy over the circular cross-section:

$$\begin{aligned} E^{\text{stacking}}(\Phi_t) &= 2\int_{-\frac{\pi}{2}}^{\frac{\pi}{2}} \int_0^R \varepsilon^{\text{stacking}}(r, \theta, \Phi_t) r\, dr d\theta \\ &= \frac{\pi(\alpha+\beta)R^4}{4}\Phi_t^2 - \frac{8\beta\delta_0 R^3}{3}\Phi_t + \pi\beta(\delta_0 R)^2 \end{aligned} \tag{4}$$

In this expression, the negative linear term indicates that twisting by an angle $\Phi_t$ lowers the stacking energy, thereby stabilizing the twisted structure. Minimizing the total energy by setting $dE^{stacking}/d\Phi_t = 0$ gives the equilibrium twisting angle as a function of the wire radius $R$.

$$\Phi_{\mathrm{t}}^{eq} = \frac{16\beta\delta_0}{3\pi R(\alpha+\beta)} \tag{5}$$

The spatial distribution of $\Delta\varepsilon^{\mathrm{stacking}} = \varepsilon^{\mathrm{stacking}}(r,\theta,\Phi_{\mathrm{t}}^{eq}) - \varepsilon^{\mathrm{stacking}}(r,\theta,0)$ at the equilibrium twisting angle clearly shows that the reduction in interlayer stacking energy away from the twin boundary is the primary driving mechanism for twisting, as shown in Figure 3i.

The twist rate, which can be obtained by dividing $\Phi_t^{eq}$ from Equation (5) by the axial distance corresponding to the two-layer stacking unit (13.193 Å), captures well the experimentally observed radius-dependent trend, as shown by the dashed line in Figure 3j. We note that, owing to the inverted-V-stacking of the layers, twisting induces an interlayer lift-up. For example, if we assume the uniform rigid layer climb of the upper layer due to the geometric constraint, an additional vertical displacement $d_z$ should be introduced (Fig. S15). When this additional effect is taken into account, Equation (5) is slightly modified with the constant $\gamma$ as

$$\Phi_{\mathrm{t}}^{eq} = \frac{16\beta\delta_0}{3\pi R(\alpha+\beta+\gamma)} \tag{6}$$

which gives better agreement with experiment as shown in Figure 3j. Here, $\gamma = 11.19\ \mathrm{meV/(f.u.\cdot Å^2)}$ is a constant related to the lift-up effect and the details are provided in the Supporting Note 1. The predicted analytical relationship, $\Phi_{\mathrm{t}}^{eq} \propto 1/R$, explains why nanowires with smaller radii exhibit larger twist rates.

We propose a generalized structural evolution mechanism in which low-symmetry layered unit cells develop into twisted structures during growth through the formation of a rotational twin boundary (Fig. 4a). Spontaneous twisting requires two essential conditions: a geometrical axis that permits rotation and a structural force imbalance that drives rotational deformation. In the present nanowires, both conditions are generated by the rotational twin boundary.

The rotational twin connects two crystalline domains with a relative rotation, producing a kink-like geometry along the growth direction. This geometry defines the twisting axis of the nanowire. As shown by the stacking-energy analysis above, the preferred interlayer registry involves a finite lateral shift, but the twin boundary constrains the corresponding interlayer sliding on opposite sides of the interface. Because regions farther from the boundary are less constrained, asymmetric interlayer relaxation develops across the nanowire. This imbalance generates a twisting torque and drives continuous lattice rotation during growth. By contrast, such torque generation is not expected for a reflection twin boundary, where symmetry prevents the formation of opposing sliding directions. Thus, the rotational twin boundary is not merely a defect but a structural element that simultaneously defines the twisting axis and converts energetically preferred interlayer sliding into rotational deformation.

This mechanism is not limited to $GeSe_{2-x}Te_x$. To test its generality, we applied the same structural concept to model low-symmetry layered systems. In $Sn_2Se_3$, a lower-symmetry unit cell containing a rotational twin boundary reproduced spontaneous rotational behavior after structural relaxation, even without any screw dislocation (Fig. 4b, c). Similarly, a cylindrical model of monoclinic 1T′-$MoTe_2$ containing a rotational twin boundary exhibited continuous layer rotation under structural relaxation (Fig. S16). These results suggest that rotational-twin-induced twisting can emerge as a generalizable structural pathway in low-symmetry layered materials with suitable symmetry and stacking conditions, providing a route for designing twist-engineered vdW nanostructures beyond the specific $GeSe_{2-x}Te_x$ system.

## Conclusion

In this work, we establish a structural mechanism in which rotational twin boundaries formed during VLS growth drive continuous lattice rotation in $GeSe_{2-x}Te_x$ nanowires. TEM and STEM analyses reveal that the twin boundary extends along the central region of the nanowire and that the layers adopt a sequential-stacking configuration characteristic of the triclinic unit cell. Unlike conventional Eshelby-twist nanowires, this twisting occurs without screw dislocations and arises from the interplay between the energy penalty imposed by the rotational twin boundary and the interlayer sliding favored in the sequential-stacking configuration. DFT calculations show that the rotational twin-boundary formation energy remains low and converges in the isolated boundary limit. The radius-dependent twist rate is explained by the competition between in-plane stacking-energy relaxation and the energetic penalty associated with interlayer lift-up deformation. These results suggest that rotational-twin-induced vdW twisting can serve as a generalizable structural pathway in low-symmetry layered materials with suitable symmetry and stacking conditions, providing a route to spontaneous chirality and twist-engineered vdW architectures.

## Methods

**$GeSe_{2-x}Te_x$ nanowire growth** $GeSe_{2-x}Te_x$ nanowires were synthesized in a two-zone quartz tube reactor using Au-catalyzed vapor–liquid–solid growth. GeSeTe powder (5–40 mg, 2D Semiconductors) was placed in the source zone and heated to 550 °C. Si(100) substrates coated with a 2.5-nm-thick Au film deposited by thermal evaporation were placed in the growth zone and heated to 370 °C for 30 min (Fig. S1a,b). During growth, Ar and $H_2$ flows were maintained at 195 and 5 sccm, respectively, under atmospheric pressure.

**Characterizations** Optical microscopy images were acquired using a Leica DM-750M. SEM images were obtained using a JEOL IT-500HR operated at 10 or 15 kV. TEM images and selected-area electron diffraction patterns were collected using a JEOL 2100Plus operated at 200 kV. Atomic-resolution STEM images were acquired using an aberration-corrected JEOL ARM200F operated at 80 or 200 kV. Energy-dispersive X-ray spectroscopy (EDS) measurements were performed using an EDS detector equipped on the JEOL

ARM200F to examine the Ge, Se, and Te composition of the synthesized nanowires. Raman spectra and mapping data were obtained using an XPER-RAM-C system. Cathodoluminescence panchromatic images and spectra were acquired using a Gatan Monarc detector attached to a Hitachi SU7000 SEM.

**Unit-cell refinement and image simulations** Unit-cell refinement was performed using multi-zone SAED patterns acquired along the nanowire axis. Radial spot positions, relative rotational angles, d-spacings, and reciprocal-vector orientations were extracted from each pattern. Reported β-$GeSe_2$ lattice parameters were used as the initial model, and the lattice constants and angles were iteratively refined to minimize the mismatch between experimental and simulated diffraction patterns. Diffraction-spot analysis was performed using Gatan DigitalMicrograph, while CrystalMaker and SingleCrystal were used for structural visualization and SAED simulation. Simulations were performed at 200 keV, with a camera length of 100.0 cm, convergence angle of 17 mrad, and sample thickness of 100 Å. Atomic-resolution STEM images were simulated using COMPUTEM with Gaussian blurring, and image contrast was enhanced using ImageJ.

**Density Functional Theory (DFT) calculations** The structural models of bulk β-$GeSe_2$, SS-$GeSe_2$, and the supercells incorporating two rotational twin boundaries were calculated using the FHI-aims package [36-38], an all-electron electronic structure method based on numeric atom-centered orbitals. For all calculations, the *tight* default basis setting was adopted. Exchange-correlation effects were described within the generalized gradient approximation (GGA) using the Perdew-Burke-Ernzerhof (PBE) functional [39]. Dispersion interactions were additionally taken into account through the Tkatchenko-Scheffler van der Waals correction scheme [40]. The Brillouin-zone sampling was determined after convergence testing. For bulk β-$GeSe_2$ and SS-$GeSe_2$, Γ-centered k-point meshes of 6 × 3 × 4 and 7 × 3 × 7 were employed, respectively. Supercells containing two rotational twin boundaries were constructed by increasing the number of repeating units from one to three. The corresponding k-point grids for these supercells were set to 7 × 7 × 2, 7 × 7 × 1, and 7 × 7 × 1. Scalar relativistic effects were included using the zero-order regular approximation (ZORA) as implemented in the FHI-aims code [41].

**Structural relaxation simulation** Structural relaxation of twinned $GeSe_2$ nanowire models was performed using the Atomic Simulation Environment (ASE) [42] with a universal machine-learning interatomic potential based on the MACE-MPA-0 model [43]. Geometry optimization was carried out using the Broyden-Fletcher-Goldfarb-Shanno (BFGS) algorithm [44-47] until the maximum atomic force on each atom was below 0.001 eV/Å. Nanowire models with radii of 16 and 20 Å were constructed, and a vacuum spacing of 15 Å was introduced along all three spatial directions to avoid interactions between periodic images.

**Acknowledgements**

This work was supported by the National Research Foundation of Korea (NRF), funded by the Korea government (MSIT) (RS-2025-00560649), the Nano & Material Technology Development Program of the NRF (RS-2024-00468995), NRF grant funded by the Korea government (MOE) (No. RS-2026-25570481), the Global Learning & Academic Research Institution for Masters, PhD students, and Postdocs (G-LAMP) Program of the NRF, funded by the Ministry of Education (No. RS-2024-00442483), and K-CHIPS (Korea Collaborative & High-tech Initiative for Prospective Semiconductor Research) funded by the Ministry of Trade, Industry and Resources (MOTIR, Korea) (No. RS-2026-25521881). Y.L. acknowledges support from the National Research Foundation of Korea (NRF) grant funded by the Korea government (MSIT) (No. RS-2026-25468648).

**Author contributions**

D.K., K. Kang, and H. K. contributed equally to this work. D.K., H.K., and M.K. synthesized the nanowires and performed Raman and cathodoluminescence measurements. D.K., H.K., K.L., Y.L., J.P., and M.J. performed TEM/STEM measurements and structural analysis. K. Kang and A.S. performed the DFT calculations. J.-E.J. provided guidance on nanowire synthesis. D.K., K. Kang, A.S. and Kwanpyo Kim wrote the manuscript with input from all authors. A.S. and K. Kim supervised the overall research.

**Competing interests**

The authors declare no other competing interests.

**Data availability**

The data supporting the findings of this study are available from the corresponding authors upon reasonable request.

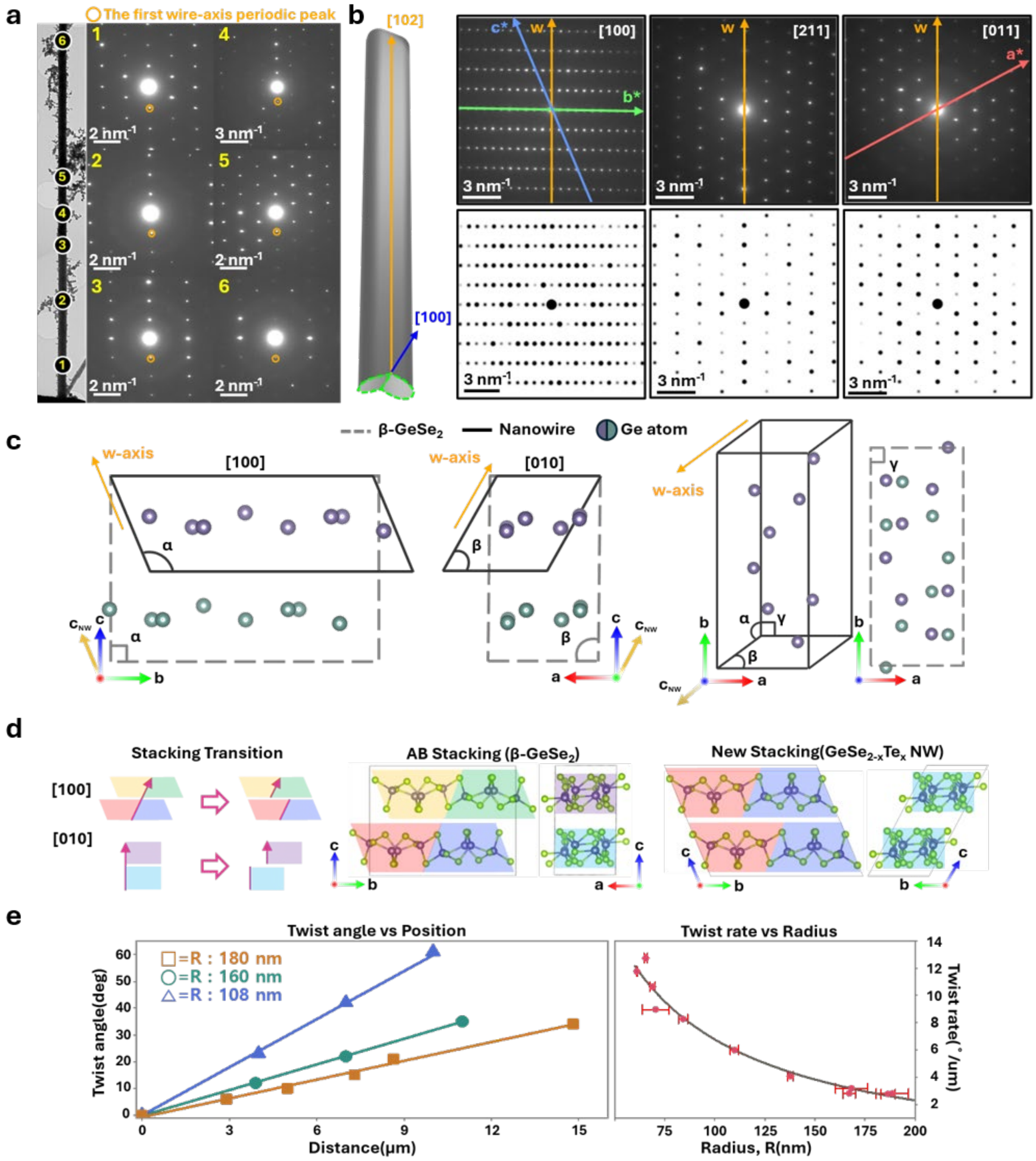

**Figure 1. Continuous twisting behavior and unit-cell/stacking reconstruction of synthesized $GeSe_{2-x}Te_x$ nanowires. a,** TEM images of a synthesized nanowire and position-resolved SAED patterns acquired along the nanowire axis. The diffraction spots associated with the growth axis, marked by orange circles, remain fixed, whereas the surrounding reciprocal-lattice spots rotate with axial position, indicating continuous lattice rotation rather than local random deformation. **b,** Schematic representation of the zone geometries observed along the nanowire, together with the corresponding low-index diffraction patterns. The orange guideline indicates the nanowire growth axis. Comparison between experimental and simulated

diffraction patterns enables indexing of the low-index zones and establishes the structural reference frame for unit-cell refinement. **c,** Refined triclinic unit cell of the synthesized nanowire, compared with the reported monoclinic β-$GeSe_2$ unit cell. Refined unitcell parameters are summarized in Table S1. **d,** Schematic comparison of the stacking sequences in β-$GeSe_2$ and the synthesized nanowire. β-$GeSe_2$ exhibits AB-type stacking, whereas the nanowire adopts a sequential-stacking arrangement, consistent with the triclinic unit cell. **e,** Twist angle as a function of axial position for nanowires with different radii, and the corresponding twist rate plotted against nanowire radius. Nanowires with smaller radii show larger twist rates, demonstrating radius-dependent continuous twisting.

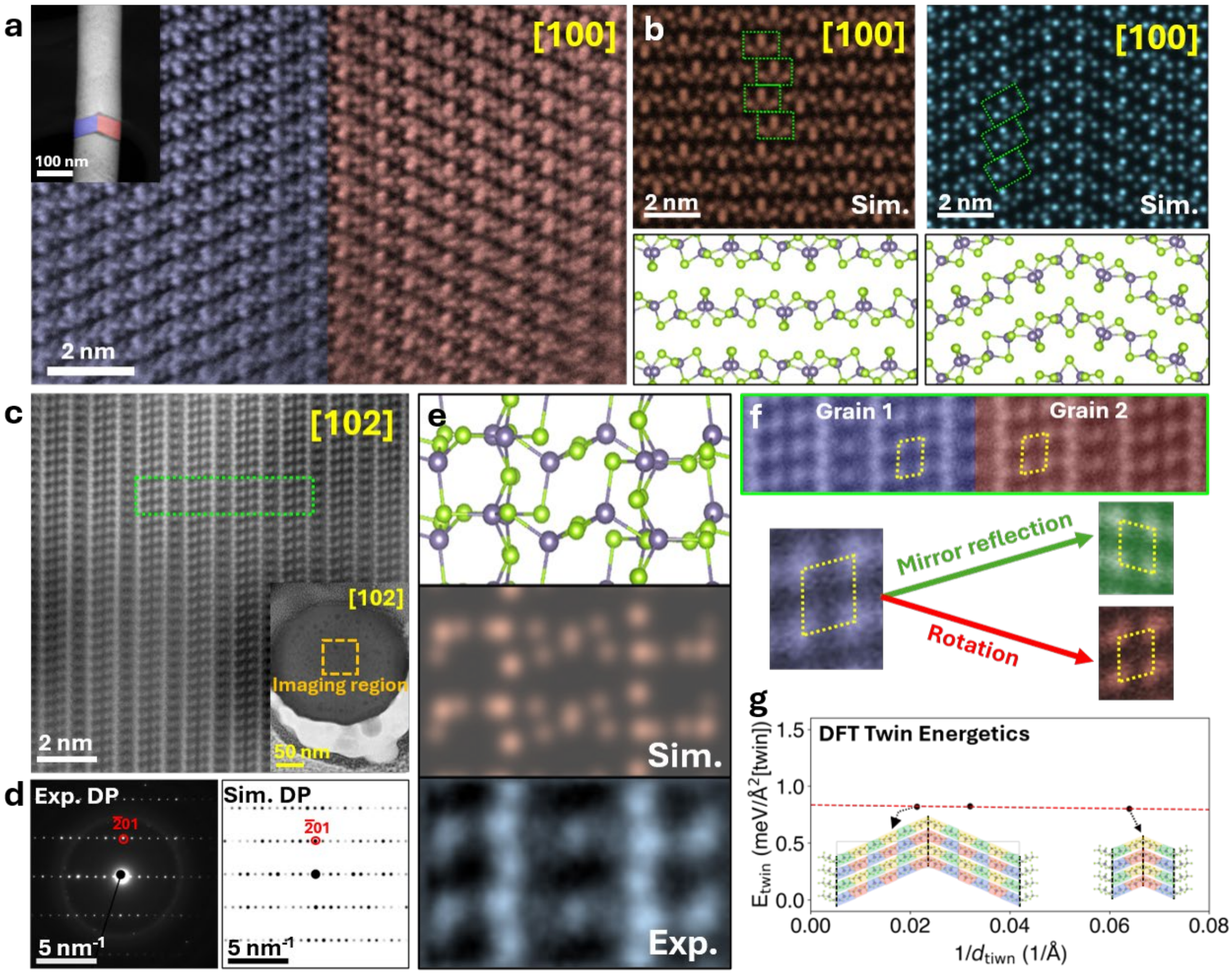

**Figure 2. Atomic-scale analysis of the rotational twin boundary. a,** Atomic-resolution STEM image acquired near the kinked region of a synthesized nanowire, indexed along the [100] zone axis. A well-defined twin boundary separates two crystalline domains, denoted as grain 1 and grain 2. **b,** Simulated STEM images of monoclinic β-$GeSe_2$ and the synthesized nanowire along the [100] zone axis. Both structures share similar Ge–Se cluster motifs, but β-$GeSe_2$ exhibits a zigzag arrangement associated with AB-type stacking, whereas the nanowire shows a sequential-stacking configuration across the twin boundary. **c,** Cross-sectional atomic-resolution STEM image obtained near the twin-boundary region of the nanowire. **d,** Comparison between the experimental diffraction pattern and the simulated diffraction pattern calculated from the refined triclinic unit cell along the [102] zone axis. The marked (−201) reflection serves as a reference, confirming the zone-axis assignment. **e,** Comparison of the refined structural model, simulated STEM image, and experimental STEM image along the [102] zone axis. **f,** Magnified views of Ge–Se cluster motifs in the two grains, with a comparison of mirror-related (green) and rotation-related (red) configurations. The motif in grain 2 matches the rotation-related configuration, supporting the assignment of a rotational

twin boundary. **g,** DFT-calculated twin-boundary formation energy as a function of inverse twin-boundary separation. The energy converges to a low value in the isolated-boundary limit, indicating that the rotational twin boundary is energetically stable.

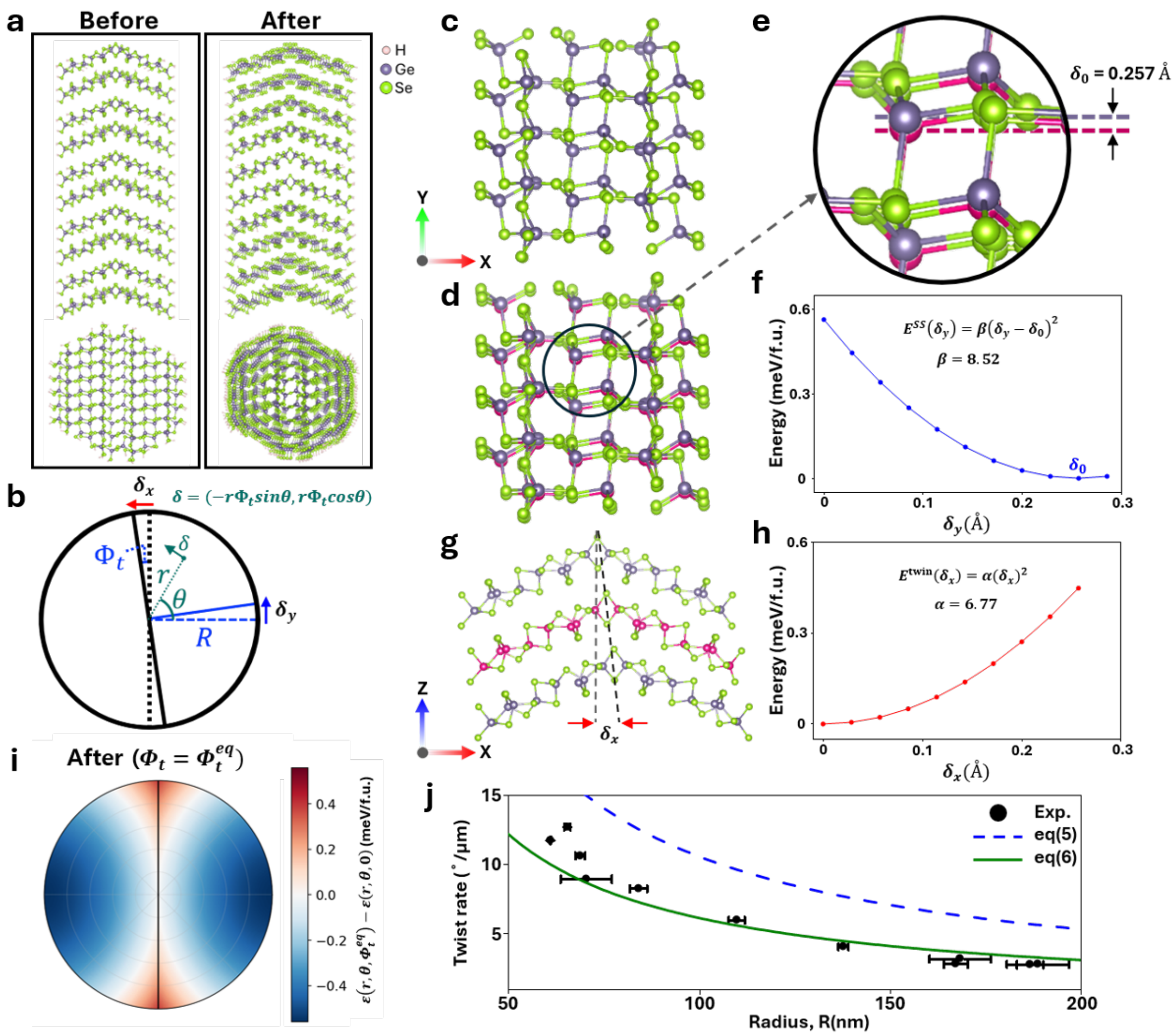

**Figure 3. Twisting mechanism driven by interlayer stacking-energy relaxation. a,** Structural relaxation simulation of an initially untwisted nanowire containing a rotational twin boundary. Side and top views before and after relaxation show the spontaneous development of continuous twisting without a screw dislocation. **b,** Geometrical model describing the local in-plane displacement induced by twisting. **c,d,** Atomic structural models showing the local stacking changes and relative displacement between adjacent layers induced by interlayer rotation. Ge atoms are represented by purple and red spheres for distinction purposes. **e,** Enlarged atomic structure showing the local interlayer displacement toward the energetically preferred sequential-stacking configuration. **f,** Energy change as a function of interlayer displacement in the direction parallel to the twin boundary ($\delta_y$) from DFT calculations. The energy is minimized at the preferred displacement $\delta_0$. **g,** Atomic structural model showing the relative displacement that induces local interlayer misalignment near the rotational twin boundary. **h,** Energy change as a function of interlayer displacement in the direction

perpendicular to the twin boundary ($\delta_x$) from DFT calculations. **i,** Cross-sectional distribution of the local stacking-energy change at the equilibrium twist angle compared with the untwisted case. **j,** Experimental twist rates as a function of nanowire radius compared with the analytical model predictions. The blue dashed curve considers only the in-plane rotation energy, whereas the green solid curve additionally includes the interlayer lift-up energy.

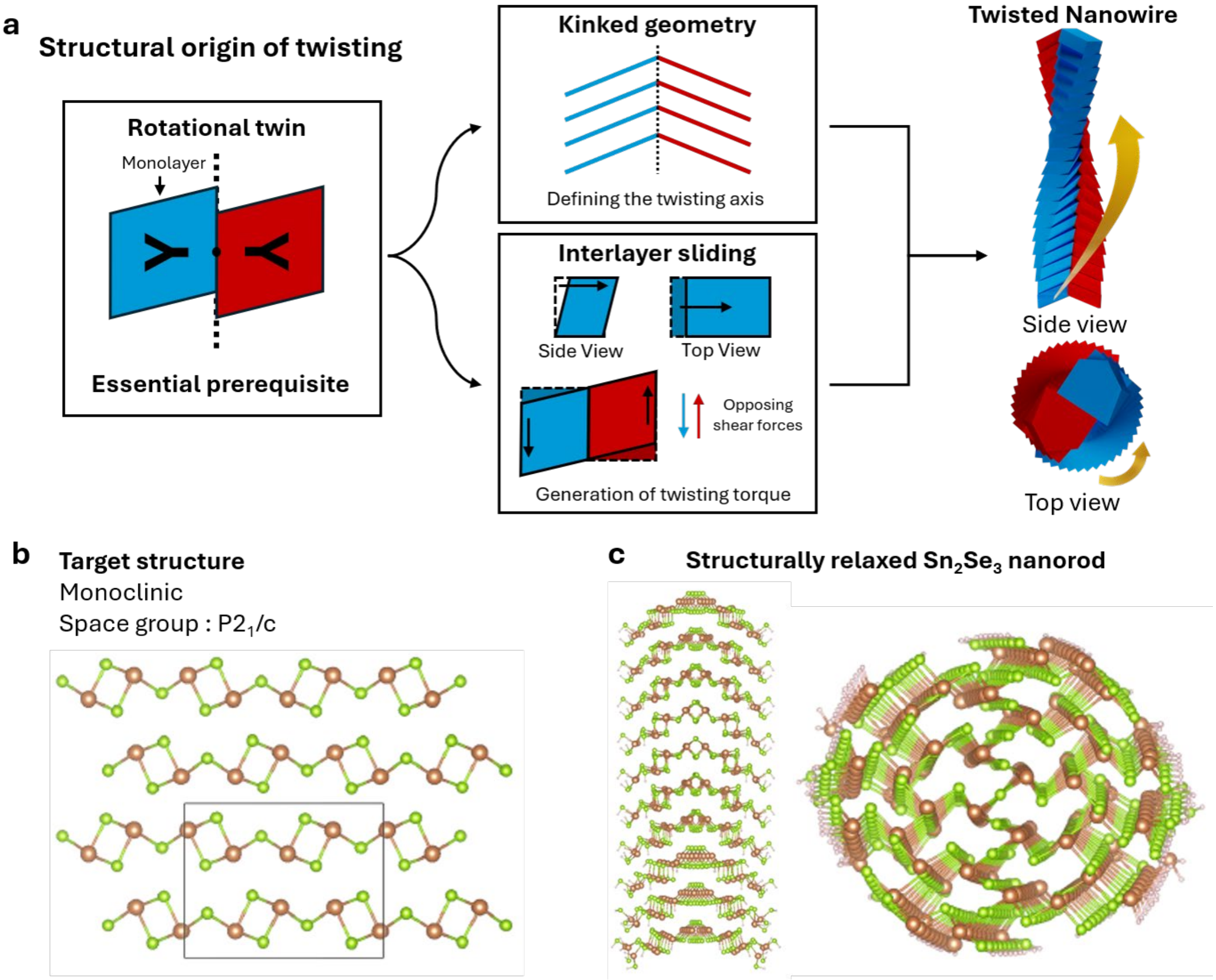

**Figure 4. Structural conditions enabling twisting in low-symmetry van der Waals materials. a,** Schematic illustration of the key structural conditions required for twisting to emerge in the nanowire. The rotational twin provides a twisting axis by forming a kinked shape and creates a structural condition in which interlayer sliding generates opposing shear forces that produce a twisting torque. When these conditions are simultaneously satisfied, the nanowire progressively rotates during growth and evolves into a continuously twisted structure. **b,** Initial structural model based on monoclinic $Sn_2Se_3$ containing an artificially introduced rotational twin. **c,** Structurally relaxed $Sn_2Se_3$ nanorod showing spontaneous twisting without a screw dislocation, demonstrating that the proposed mechanism is not specific to $GeSe_{2-x}Te_x$ nanowires.

**Supporting Information for**

# Interplay between Interlayer Shift and Twist: Twisted van der Waals Nanowires Driven by Rotational Twinning

*Dong-gyu Kim[1,†], Kisung Kang[2,3,†], Hani Kang[1,†], Kihyun Lee[1], Yangjin Lee[1,4], Jinsub Park[1], Joong-Eon Jung[1], Min Kim[1], Myeongjin Jang[1], Aloysius Soon[2,*], and Kwanpyo Kim[1,*]*

[1]Department of Physics, Yonsei University, Seoul 03722, Republic of Korea.

[2]Department of Materials Science & Engineering, Yonsei University, Seoul 03722, Republic of Korea

[3]School of Materials Science and Engineering, Chonnam National University, Gwangju 61186, Republic of Korea

[4]Institute of Advanced Composite Materials, Korea Institute of Science and Technology (KIST), Jeonbuk 55324, Korea

*Corresponding Authors: A.S. (aloysius.soon@yonsei.ac.kr), K.K. (kpkim@yonsei.ac.kr)

[†]Equal Contributions

**Supporting Note 1. Interlayer lift-up deformation and modified equilibrium twist angle**

In addition to the in-plane stacking-energy relaxation described in the main text, relative rotation between the inverted-V-stacked layers induces an additional out-of-plane deformation. For an interlayer rotation angle $\Phi_t$ associated with one two-layer stacking unit, the in-plane displacement at the outer edge of a nanowire with radius $R$ is approximated, under the small-angle condition, as

$$\delta = R\Phi_t. \tag{S1}$$

Because of the inclined geometry of the inverted-V stacking configuration, this in-plane displacement is accompanied by an interlayer lift-up displacement $d_z$. The geometrical relation between the two displacements is expressed as

$$d_z = \delta \tan\theta_k = R\Phi_t \tan\theta_k, \tag{S2}$$

where $\theta_k$ denotes the effective inclination angle of the inverted-V stacking geometry, as illustrated in Figure S15a.

To quantify the energetic cost associated with the interlayer lift-up deformation, Density Functional Theory (DFT) calculations were performed by systematically varying $d_z$. The calculated relative energy shown in Figure S15b is well described by a quadratic function:

$$\Delta\varepsilon_{\mathrm{lift}}(d_z) = Ad_z^2, \tag{S3}$$

where $A = 15.52\ \mathrm{meV/(f.u.\cdot Å^2)}$ is the quadratic coefficient obtained from the DFT calculations.

The DFT-derived local energies used in the main text and here are normalized per formula unit. Conversion to continuum cross-sectional energies formally introduces a common areal number density of formula units, $\rho_{\mathrm{f.u.}}$. Because this factor multiplies both the in-plane stacking and lift-up energy contributions and cancels upon minimization with respect to $\Phi_t$, it is omitted for simplicity.

Using the geometrical relation in Equation (S2), the lift-up energy over the circular nanowire cross-section is expressed as

$$E_{\text{lift}}(\Phi_t) = \pi R^2 A (R\Phi_t \tan\theta_k)^2. \quad \text{(S4)}$$

Expanding Equation (S4) gives

$$E_{\text{lift}}(\Phi_t) = \pi A R^4 \tan^2\theta_k \, \Phi_t^2. \quad \text{(S5)}$$

The total rotational energy is defined as the sum of the in-plane stacking energy derived in the main text and the lift-up energy:

$$E_{\text{total}}(\Phi_t) = E_{\text{stacking}}(\Phi_t) + E_{\text{lift}}(\Phi_t). \quad \text{(S6)}$$

Using the stacking-energy expression derived in the main text, the total energy is written as

$$E_{\text{total}}(\Phi_t) = \pi R^4 \left[\frac{\alpha + \beta}{4} + A\tan^2\theta_k\right] \Phi_t^2 - \frac{8\beta\delta_0 R^3}{3} \Phi_t + \pi\beta\delta_0^2 R^2. \quad \text{(S7)}$$

The equilibrium twist angle is obtained by minimizing the total energy with respect to $\Phi_t$,

$$\frac{dE_{\text{total}}}{d\Phi_t} = 0,$$

which yields

$$\Phi_t^{\text{eq}} = \frac{16\beta\delta_0}{3\pi R[\alpha + \beta + 4A\tan^2\theta_k]}. \quad \text{(S8)}$$

For consistency with Equation (6) in the main text, the effective coefficient associated with the lift-up energy is defined as

$$\gamma \equiv 4A\tan^2\theta_k.$$

Using $A = 15.52\ \text{meV}/(\text{f.u.}\cdot\text{Å}^2)$ and $\theta_k = 23°$, this coefficient is evaluated as

$$\gamma = 4(15.52)\tan^2(23°) \approx 11.19\ \text{meV}/(\text{f.u.}\cdot\text{Å}^2),$$

where $\gamma$ has the same units as $\alpha$and $\beta$. The equilibrium twist angle can therefore be written as

$$\Phi_t^{\mathrm{eq}} = \frac{16\beta\delta_0}{3\pi R(\alpha + \beta + \gamma)}. \tag{S9}$$

Equation (S9) is identical to Equation (6) in the main text.

The lift-up deformation introduces an additional positive energy penalty during interlayer rotation and consequently reduces the equilibrium twist angle compared with the model considering only the in-plane stacking energy. Nevertheless, the inverse dependence on the nanowire radius is preserved:

$$\Phi_t^{\mathrm{eq}} \propto \frac{1}{R}.$$

Thus, the lift-up-corrected model predicts a smaller twist rate than the stacking-energy-only model while retaining the experimentally observed decrease in twist rate with increasing nanowire radius.

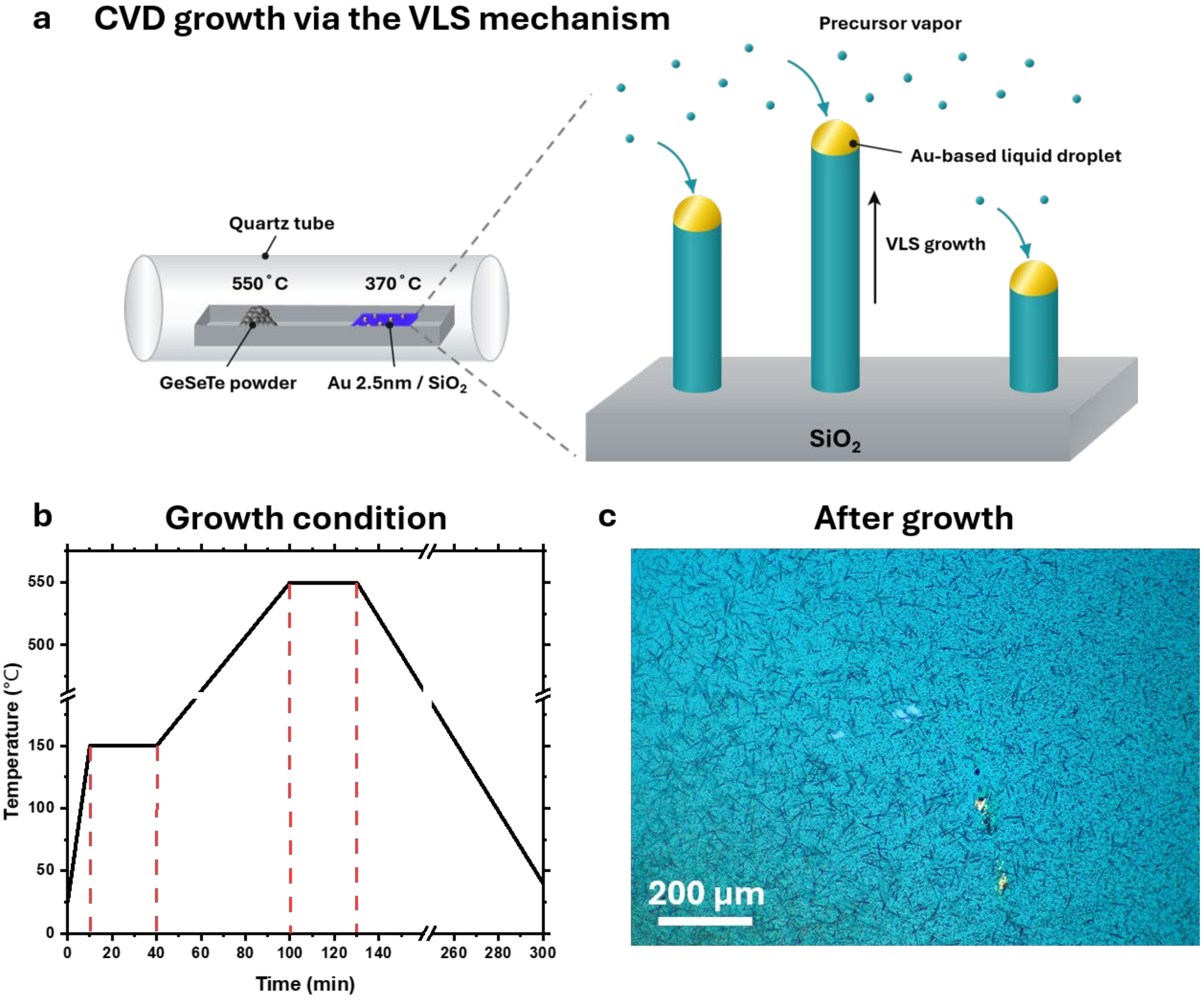

**Figure S1. Synthesis of $GeSe_{2-x}Te_x$ nanowires. a,** Schematic illustration of $GeSe_{2-x}Te_x$ nanowire growth via the VLS mechanism in a CVD system. **b,** Temperature profile for $GeSe_{2-x}Te_x$ nanowire growth. **c,** Optical microscopy image of VLS-grown $GeSe_{2-x}Te_x$ nanowires.

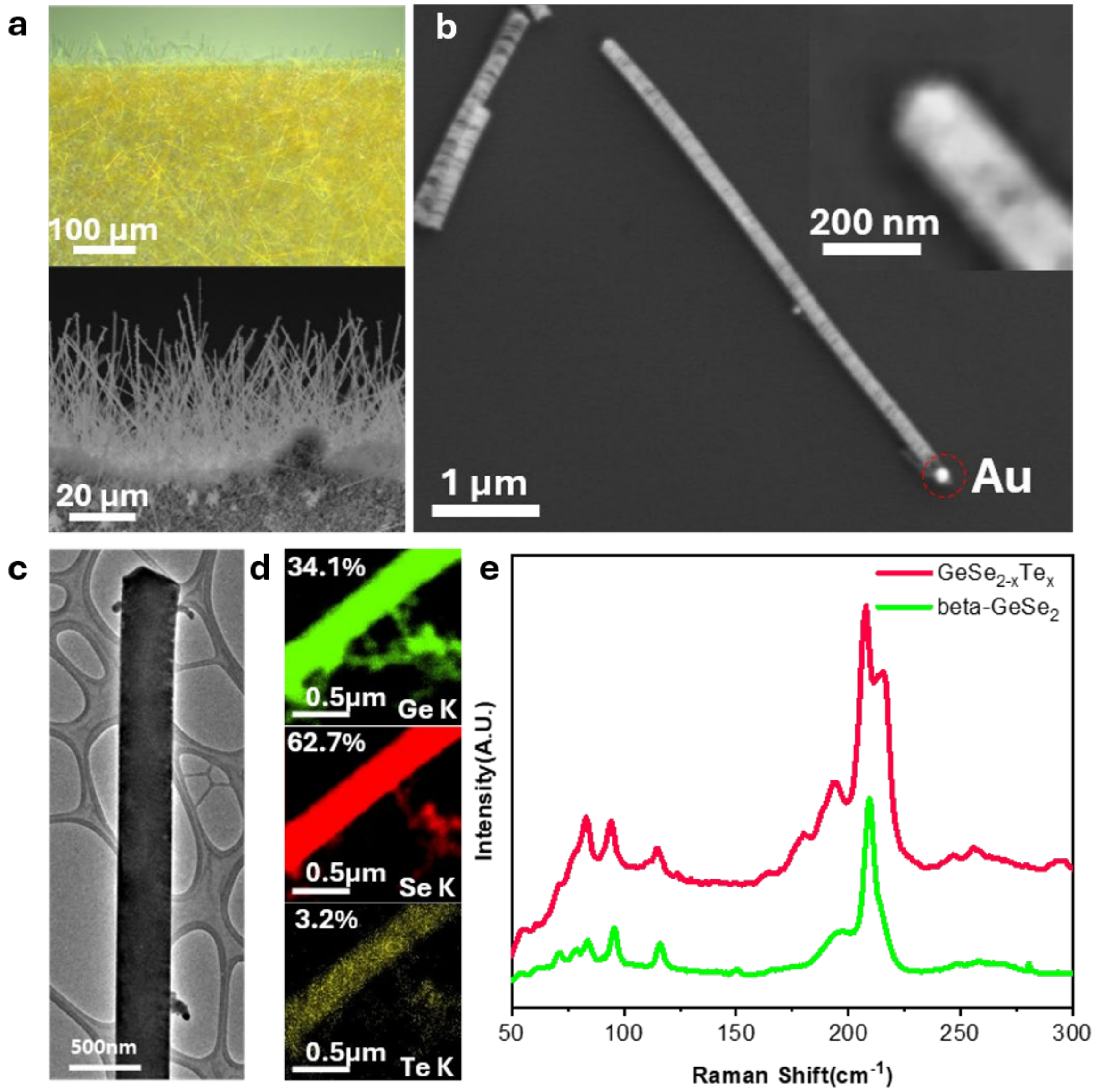

**Figure S2. Structural and compositional characterization of synthesized $GeSe_{2-x}Te_x$ nanowires. a,** Optical microscopy and SEM images of synthesized nanowires on a $SiO_2$/Si substrate. **b,** SEM image of a VLS-grown nanowire showing an Au catalyst at the tip and a kinked structure, which identifies the nanowire growth direction. **c,** TEM image of a nanowire exhibiting kinked morphology. **d,** EDS analysis showing the Ge, Se, and Te composition of the synthesized nanowires. **e,** Raman spectra of reported β-$GeSe_2$ and the synthesized nanowires.

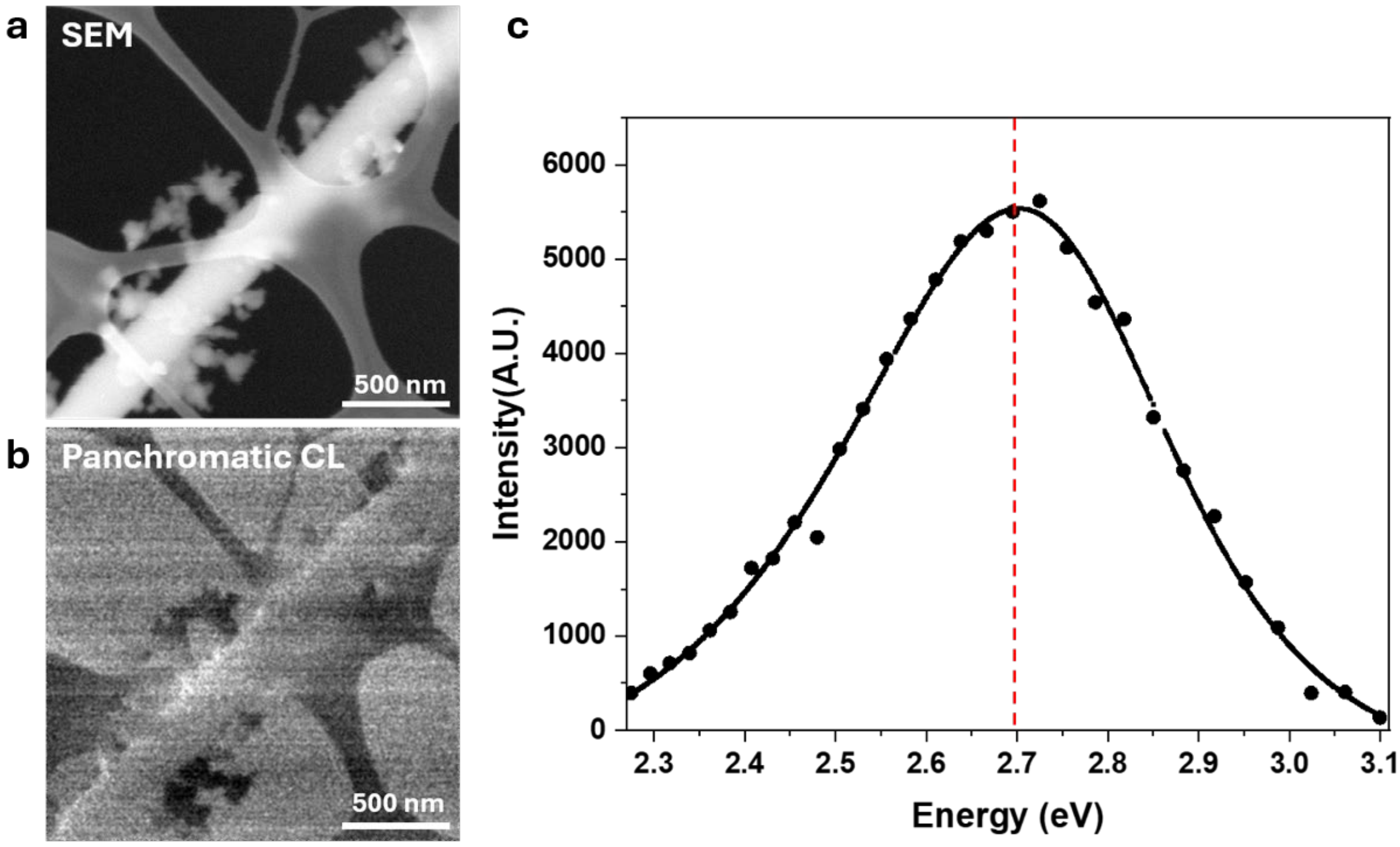

**Figure S3. Cathodoluminescence characterization of a synthesized nanowire. a,** SEM image of the synthesized nanowire. **b,** Panchromatic CL mapping image acquired from the same region. **c,** Averaged CL spectrum of the synthesized nanowire, showing a sharp emission peak at approximately 2.70 eV.

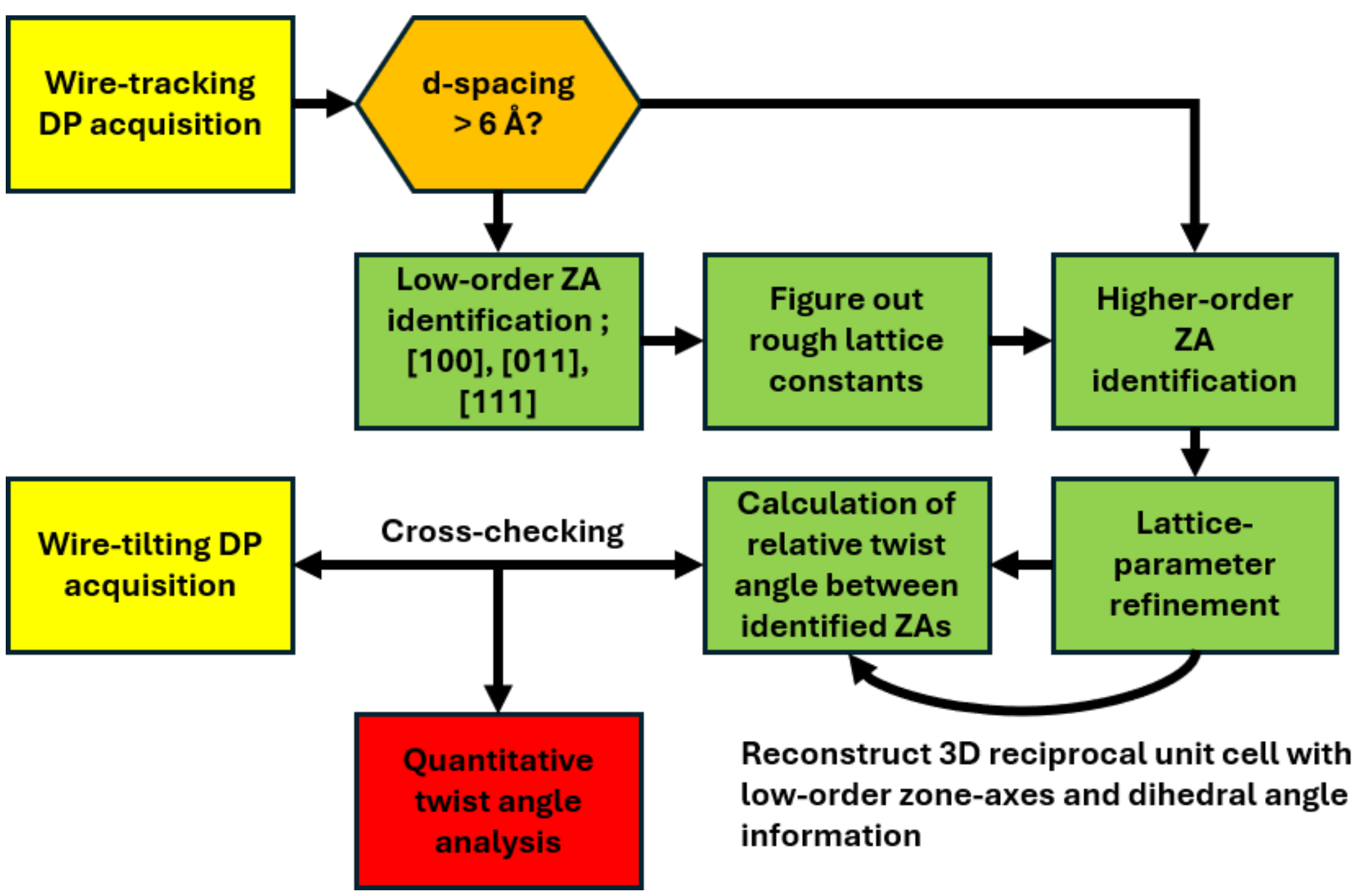

**Figure S4. Workflow for experimental unit-cell refinement and twist angle analysis using diffraction patterns acquired from nanowires.** This flowchart summarizes the step-by-step procedure used to refine the unit cell of the synthesized nanowires and to quantitatively analyze the twist angle along the wire axis based on experimentally obtained diffraction patterns. Diffraction patterns are first acquired under wire-tracking and wire-tilting conditions. Low-order zone axes are then identified to estimate the initial lattice parameters, followed by higher-order zone-axis indexing to further refine the lattice constants. The relative twist angles between successive zone axes are subsequently determined, enabling quantitative analysis of the continuous twisting behavior along the nanowire.

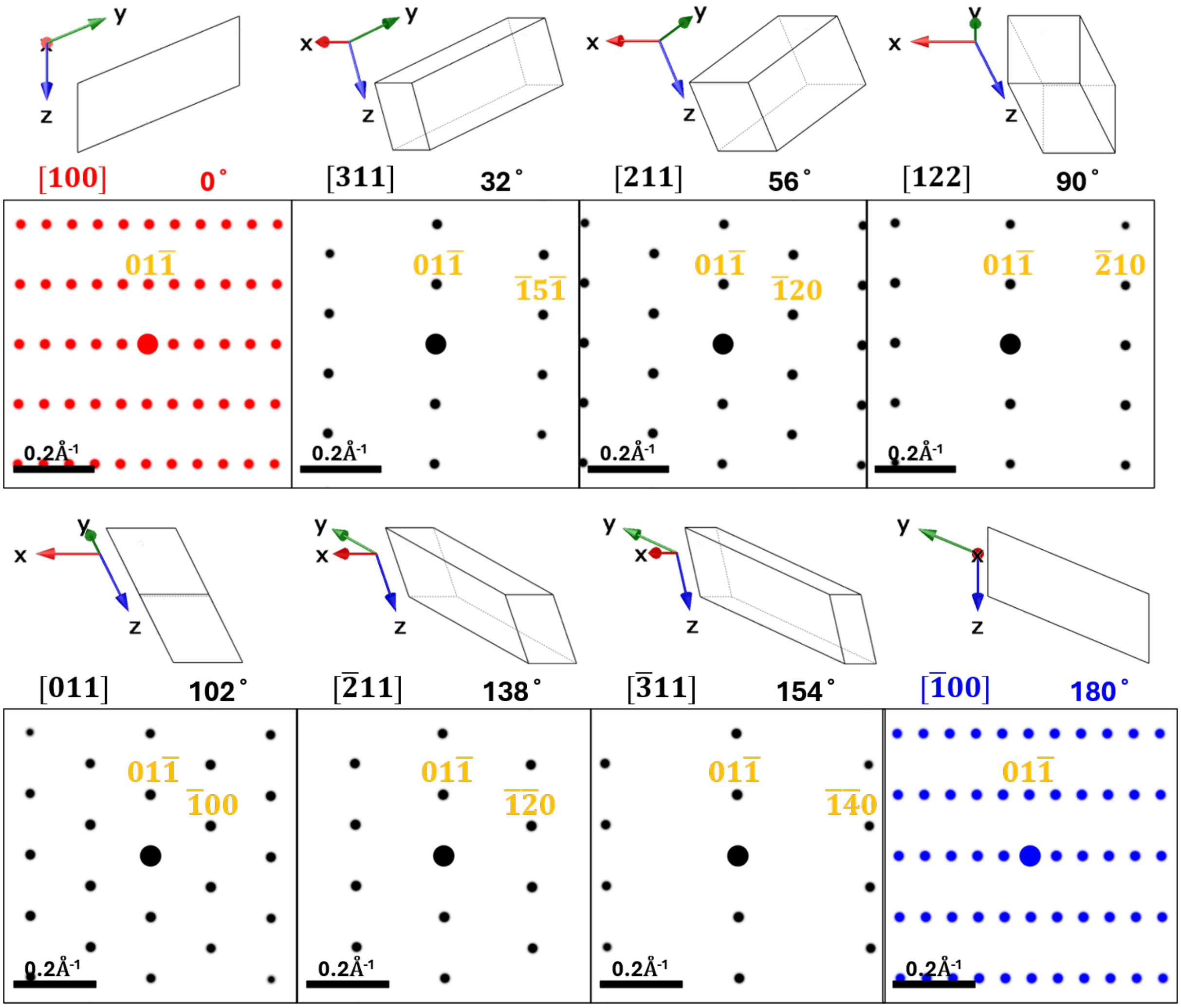

**Figure S5. Simulated diffraction-pattern evolution during 180° lattice rotation.** Simulated [100] zone-axis diffraction patterns during rotation of the defined unit cell from 0° to 180°. The unit-cell schematics indicate the corresponding lattice orientation at each rotation angle. The indexed reflections show the continuous evolution of diffraction-spot positions, and the nearly equivalent 0° and 180° patterns support the initial unit-cell definition without explicitly considering the rotational twin boundary.

**a** β-$GeSe_2$

AB Stacking (AB) Bulk : SG 14($P2_1/c$)

**b Refined nanowire structure**

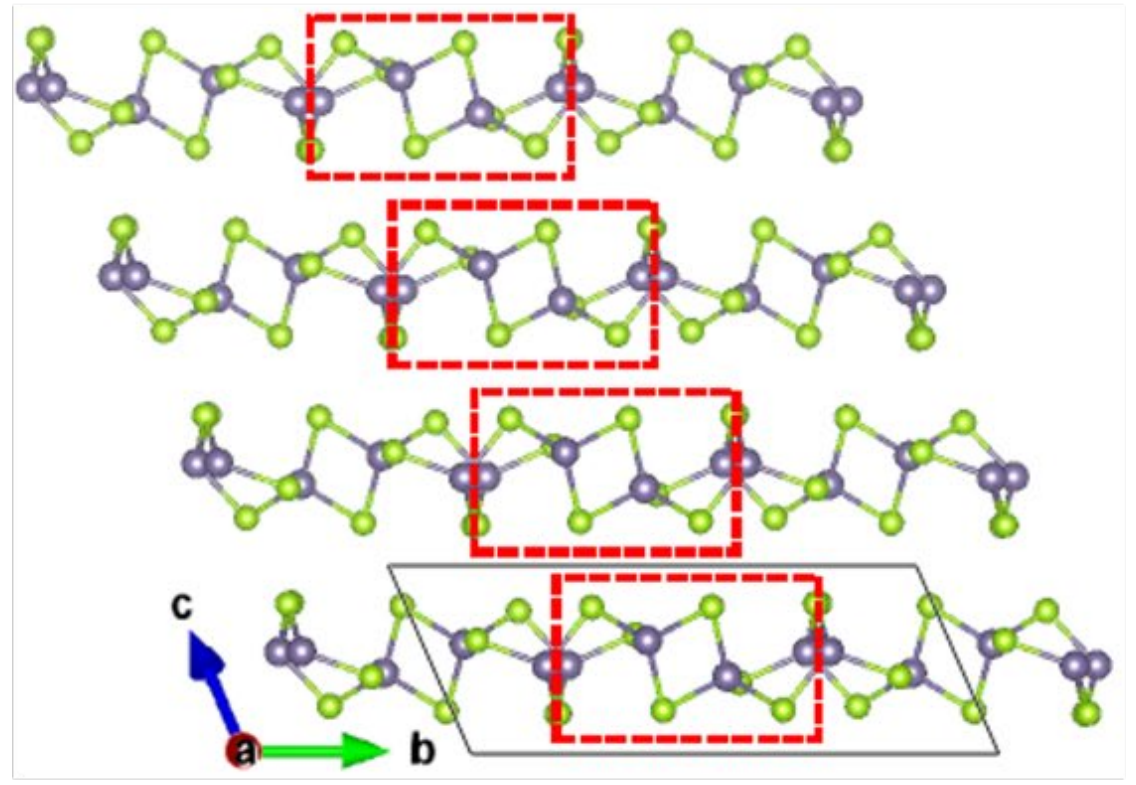

Sequential Stacking(SS) Bulk : SG 1(P1)

**c SS/AB Bulk Comparison**

$$\Delta \boldsymbol{E} = \boldsymbol{E}_{SS} - \boldsymbol{E}_{AB}$$

| [eV/f.u.] | AB Bulk | SS Bulk | $\Delta E_{SS-AB}$ |
|---|---|---|---|
| Energy | -4.259912 | -4.253026 | 0.006887 |

**Figure S6. DFT comparison between β-$GeSe_2$ and the refined sequential-stacking nanowire structure.** Structural models of β-$GeSe_2$ with AB-type stacking and the refined nanowire structure with sequential stacking. The relative energy difference, defined as $\boldsymbol{\Delta E} = \boldsymbol{E_{SS}} - \boldsymbol{E_{AB}}$, is small, indicating that the sequential-stacking configuration is energetically accessible and can be stabilized in the synthesized nanowire structure.

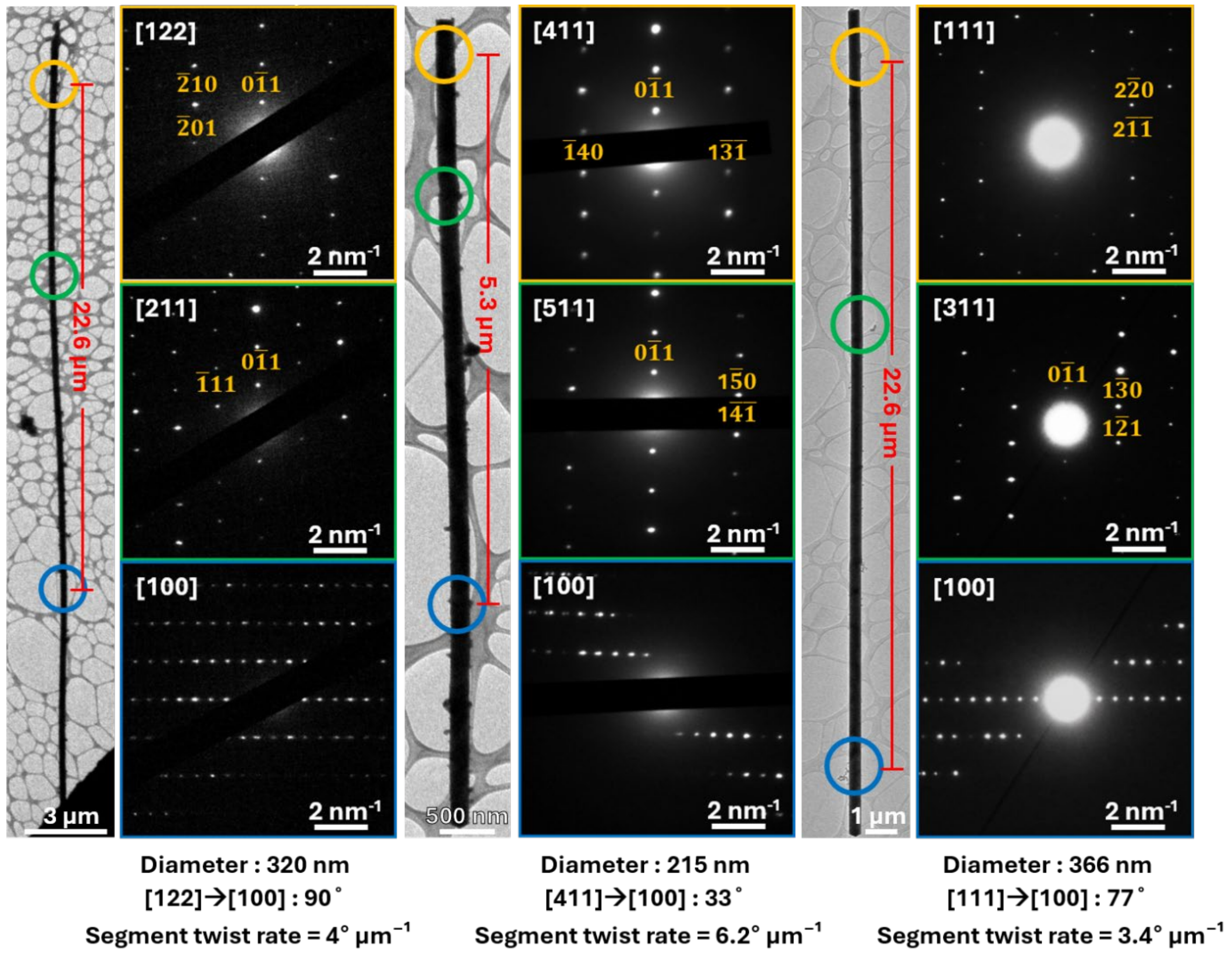

**Figure S7. Position-resolved SAED analysis supporting the diameter-dependent twist-rate estimation.** Representative TEM images and SAED patterns acquired at selected positions along synthesized $GeSe_{2-x}Te_x$ nanowires with different diameters. The colored circles indicate the positions where distinguishable zone-axis SAED patterns were obtained. The local lattice orientation at each position was identified by comparison with simulated diffraction patterns, and the twist rate was estimated from the angular change in lattice orientation along the wire axis. These additional examples further support the diameter-dependent variation in twist rate shown in Fig. 1e.

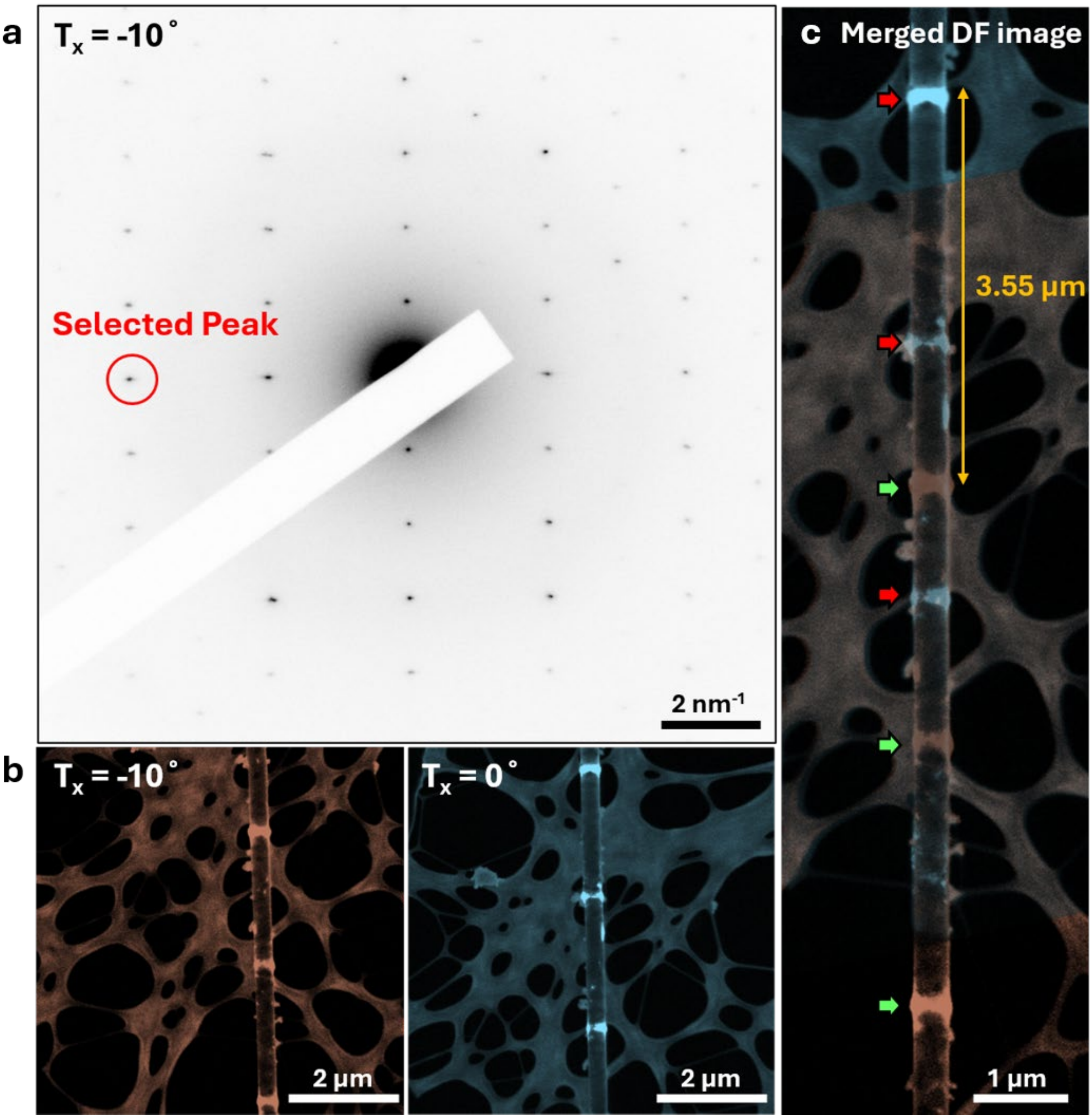

**Figure S8. Tilt-dependent dark-field TEM imaging of a synthesized nanowire. a,** Diffraction pattern acquired $T_x = -10^\circ$, with the selected peak for DF imaging marked by a red circle. **b,** DF-TEM images acquired at $T_x = -10^\circ$ and $T_x = 0^\circ$, showing different bright-contrast regions along the nanowire. **c,** Merged DF image constructed from the DF images in **b**, visualizing the shift of DF-bright regions along the nanowire upon changing the specimen tilt angle. This indicates that different positions along the nanowire satisfy the selected diffraction condition at different tilt angles.

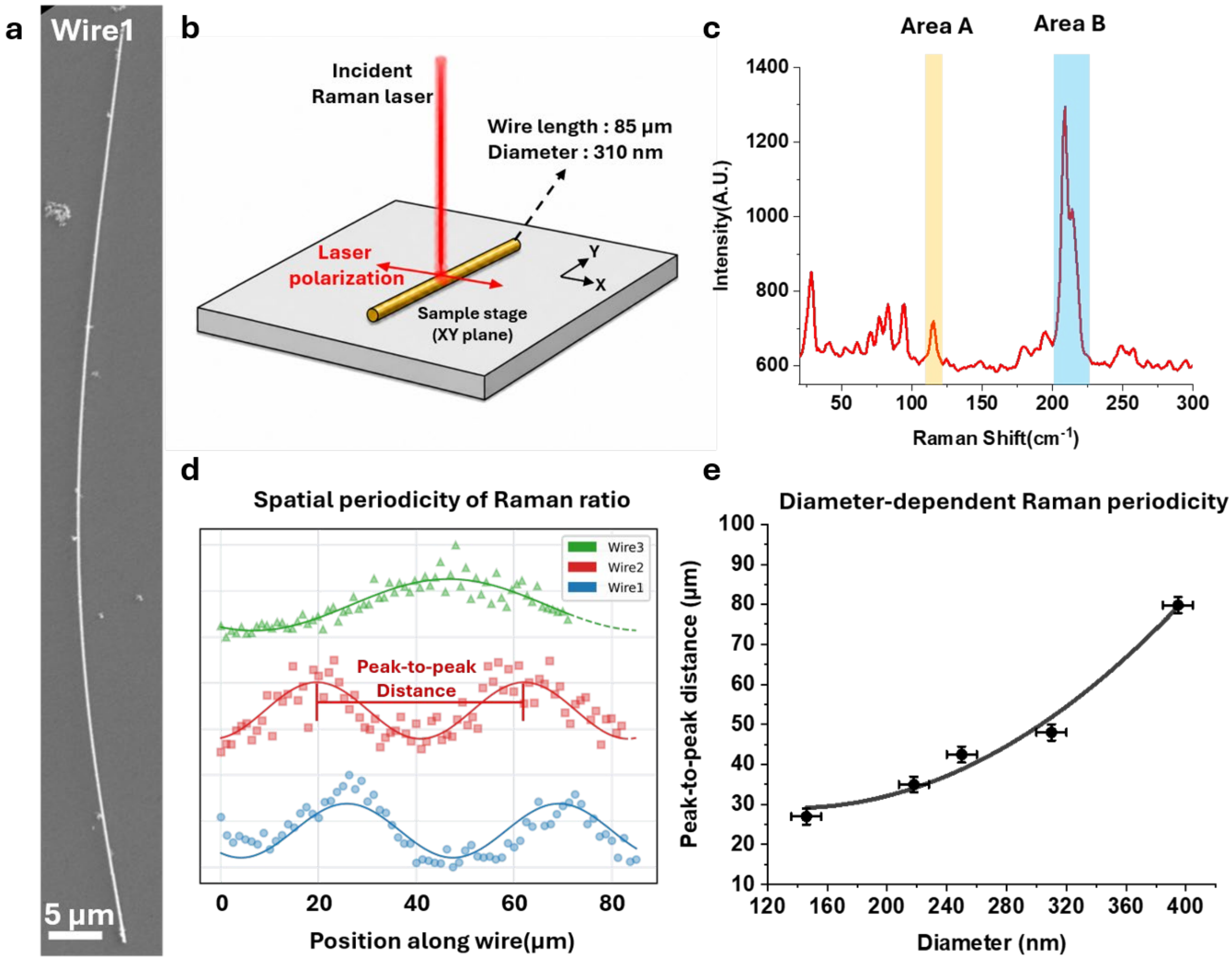

**Figure S9. Raman ratio mapping of twist periodicity in synthesized nanowires. a,** Scanning electron microscopy (SEM) image of Wire 1 used for Raman line-scan measurements. **b,** Raman measurement geometry with laser polarization perpendicular to the nanowire axis. **c,** Representative Raman spectrum highlighting Areas A and B, corresponding to the $A_g^2$ and $A_g^1$ modes, respectively. **d,** Spatial modulation of the B/A intensity ratio for three nanowires with different diameters, orientations, and twist rates. Although the phase and waveform differ among the wires, clear oscillatory behavior is observed in all cases. **e,** Peak-to-peak distance as a function of nanowire diameter, showing an increase with increasing diameter, consistent with the diffraction-derived twist-rate trend. Horizontal error bars indicate a diameter uncertainty of ±10 nm, and vertical error bars indicate an estimated uncertainty of ±2 μm in determining the peak-to-peak distance.

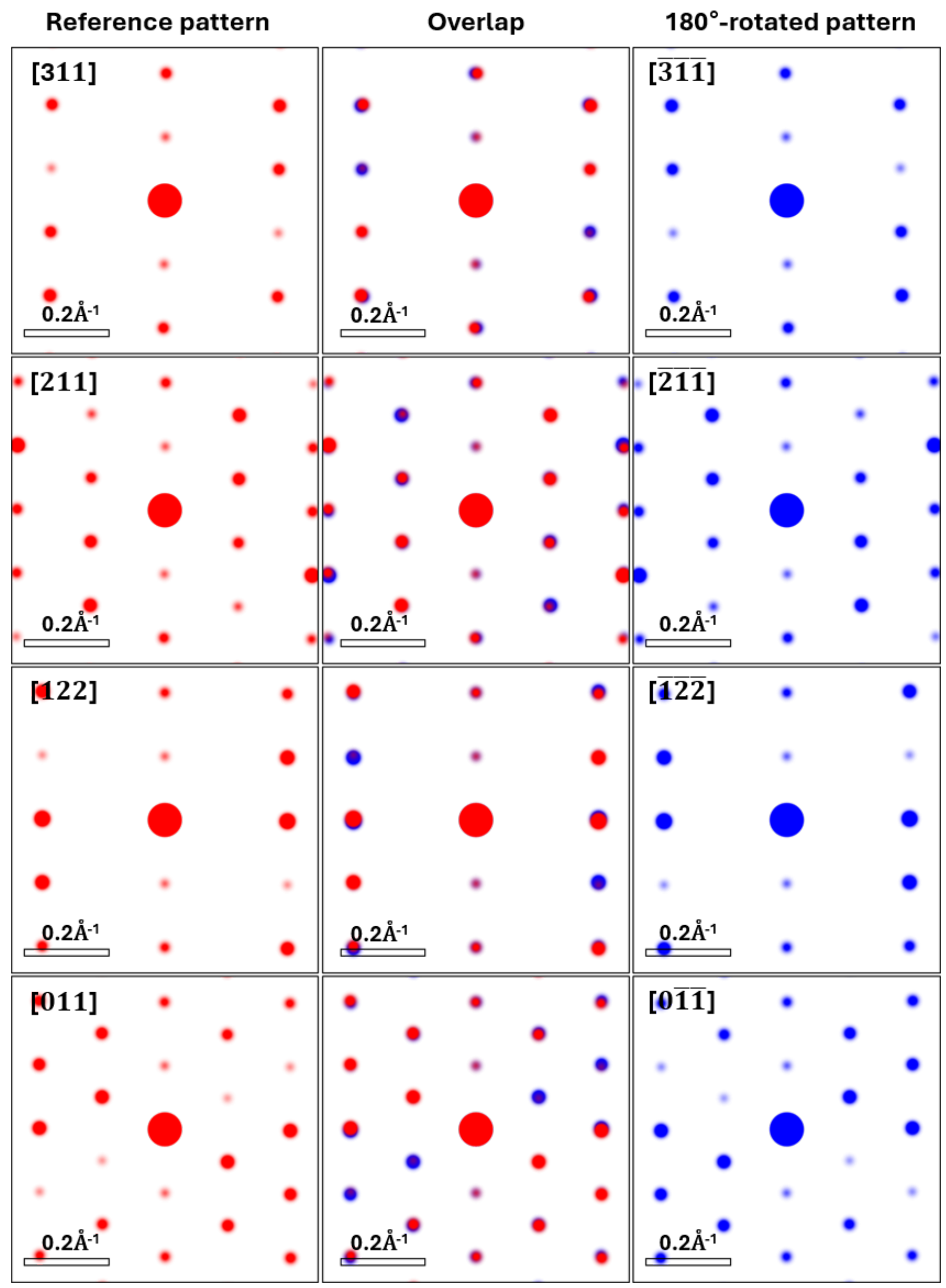

**Figure S10. Simulated diffraction-pattern overlap for twin-related unit-cell models.** Reference and 180°-rotated diffraction patterns largely overlap across multiple unit-cell models, explaining why the rotational twin boundary was not resolved by electron diffraction alone.

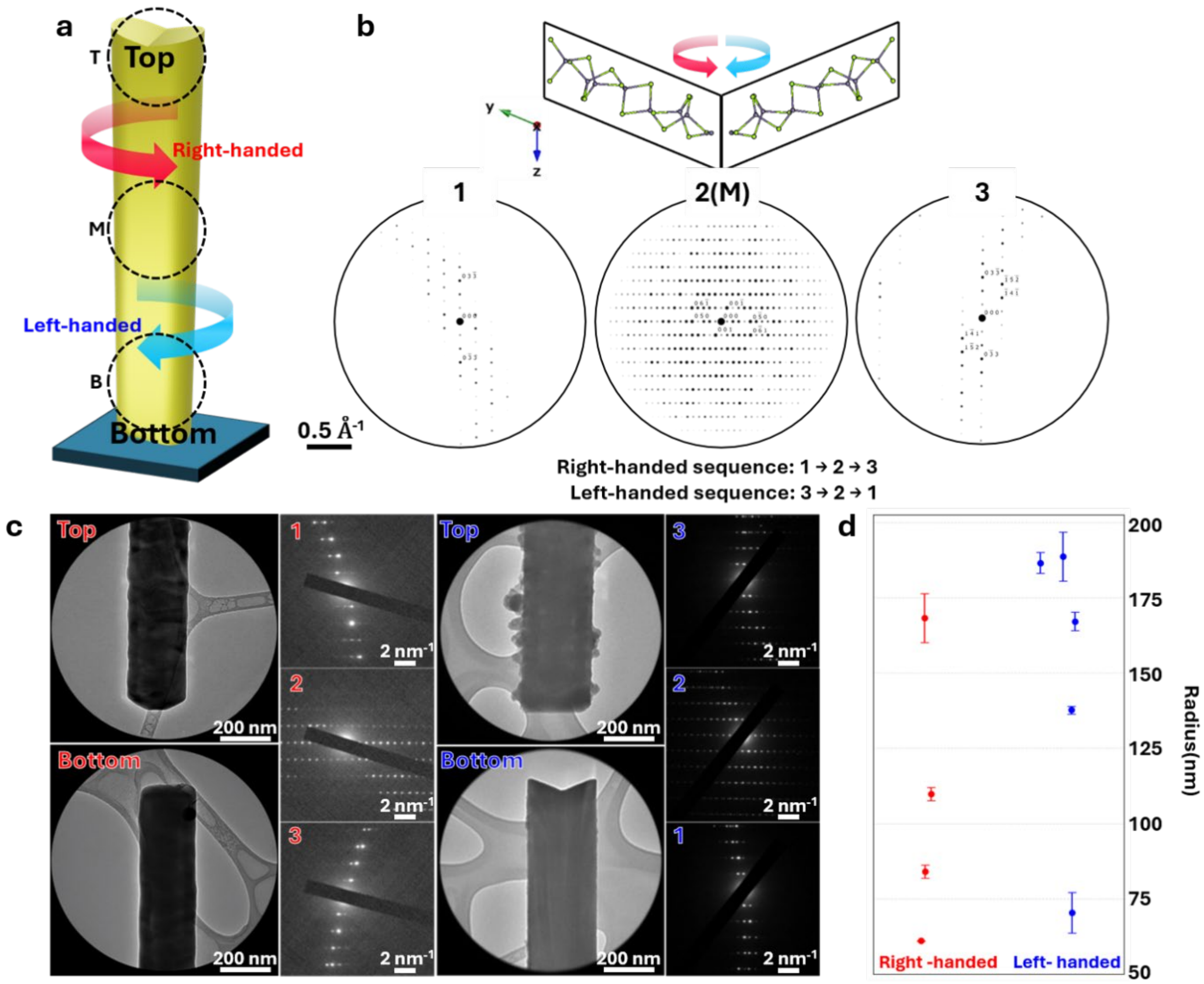

**Figure S11. Handedness analysis of synthesized nanowires. a,** Schematic defining the top/bottom regions and handedness convention. The kink-opening side was assigned as the top region. **b,** Simulated diffraction-pattern states used as a reference for handedness assignment. For the same top-to-bottom direction, right-handed nanowires follow the sequence 1 → 2 → 3, whereas left-handed nanowires follow 3 → 2 → 1. **c,** Representative right- and left-handed nanowires with corresponding TEM images and diffraction patterns. **d,** Radius distribution of nanowires with assigned handedness.

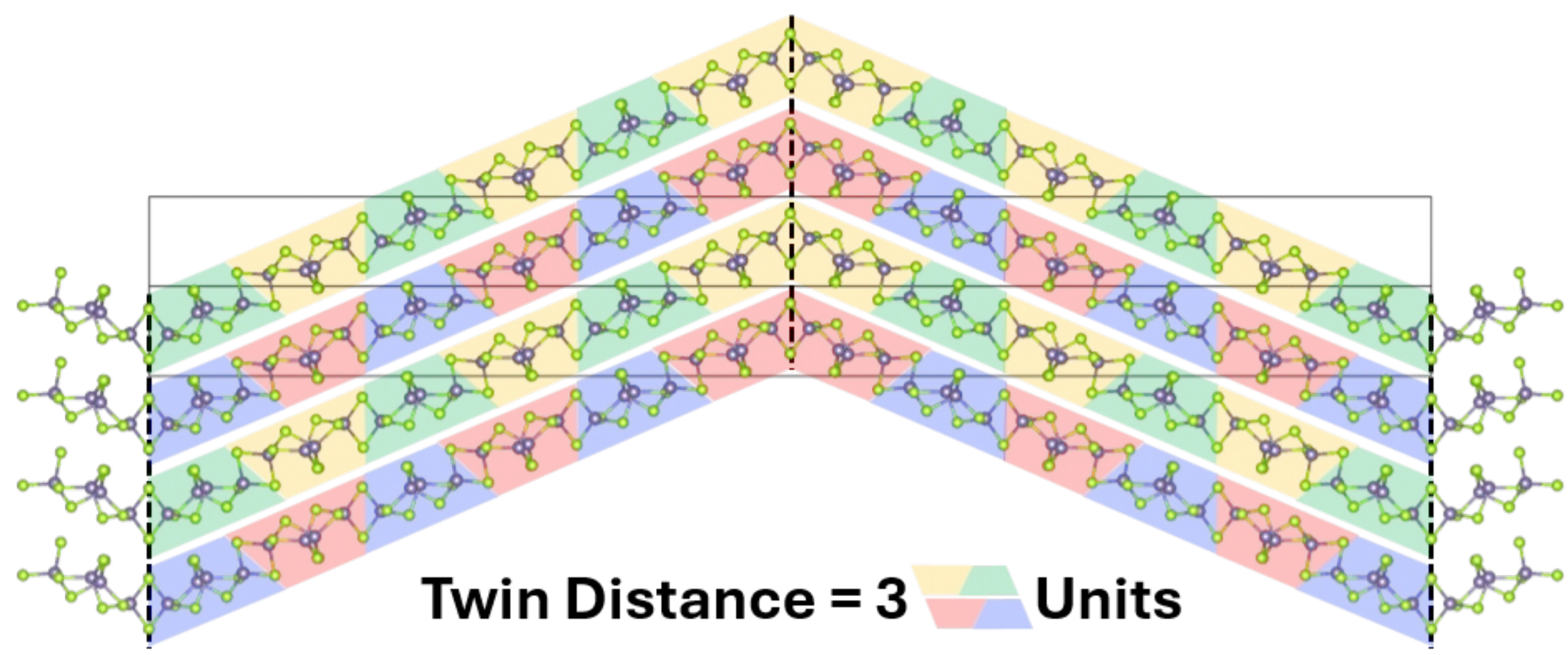

$$E_{\text{twin}} = (E_{N\ \text{Units}} - 2N \times E_{SS})/(2 \times A_{\text{twin}})$$

| ***N*** | $E_{N\ \text{Units}}$ [eV/V$_{\text{cell}}$] | $E_{SS}$ [eV/V$_{\text{cell}}$] | $A_{\text{twin}}$ [Å$^2$] | $E_{\text{twin}}$ [eV/Å$^2$] |
|---|---|---|---|---|
| **1 Unit** | -68.046 | -34.061 | 47.083 | 0.00080 |
| **2 Units** | -136.165 | -34.061 | 47.033 | 0.00082 |
| **3 Units** | -204.287 | -34.061 | 47.019 | 0.00082 |

Rotation Twin Formation Energy: $E_{\text{twin}}(d_{\text{twin}} \rightarrow \infty) = 0.00083$ eV/Å$^2$

**Figure S12. DFT model for the rotational twin-boundary formation energy calculation.** Periodic supercell model containing two rotational twin boundaries constructed from the sequential-stacking $GeSe_2$ structure. DFT calculations based on this twinned supercell give a low rotational twin-boundary formation energy, converging to approximately 0.00083 eV/Å$^2$ in the isolated-boundary limit.

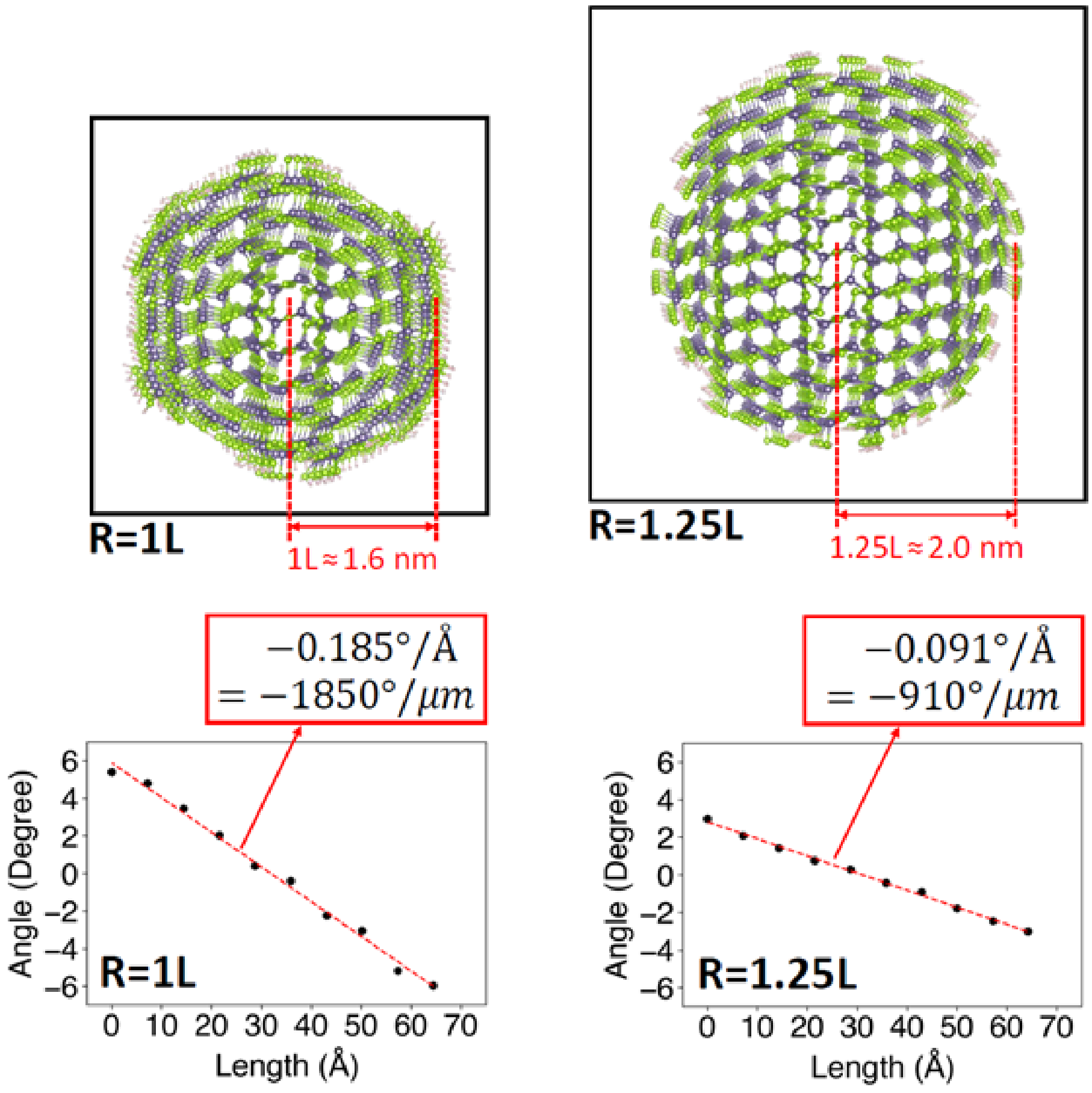

**Figure S13. Simulation verification of the radius dependence of the twist rate.** Simulation results for nanowire models with different radii (R = 1L and R = 1.25L) are presented. The top panels show three-dimensional structural models of the nanowires with effective radii of approximately 1.6 nm and 2.0 nm, respectively. The bottom panels provide a quantitative comparison of the calculated twist angle evolution for each radius. The negative slopes indicate the direction of lattice rotation under the chosen angular convention. In both cases, the twist angle exhibits linear behavior, while the magnitude of the twist rate decreases systematically with increasing nanowire radius. These results confirm, at the simulation level, the radius dependence of the twist-rate magnitude.

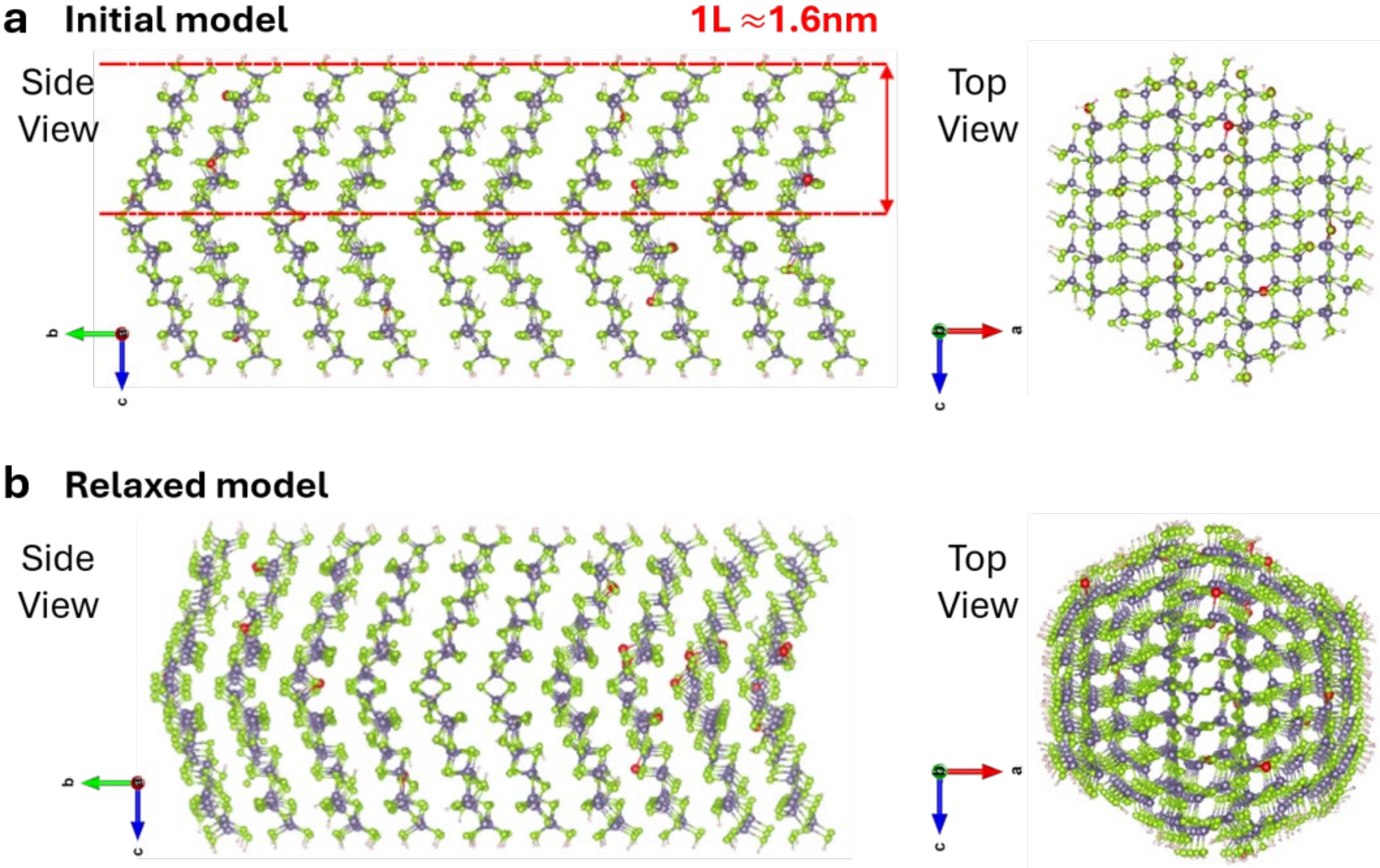

**Figure S14. Structural relaxation of the Te-incorporated twinned $GeSe_2$ nanowire model.** Initial and relaxed structures of a Te-incorporated $GeSe_2$ nanowire model containing a rotational twin boundary. The relaxed structure develops a twisted morphology, indicating that Te incorporation is not essential for inducing the twisting behavior and that the rotational-twin-based structural motif is the primary origin of the twist.

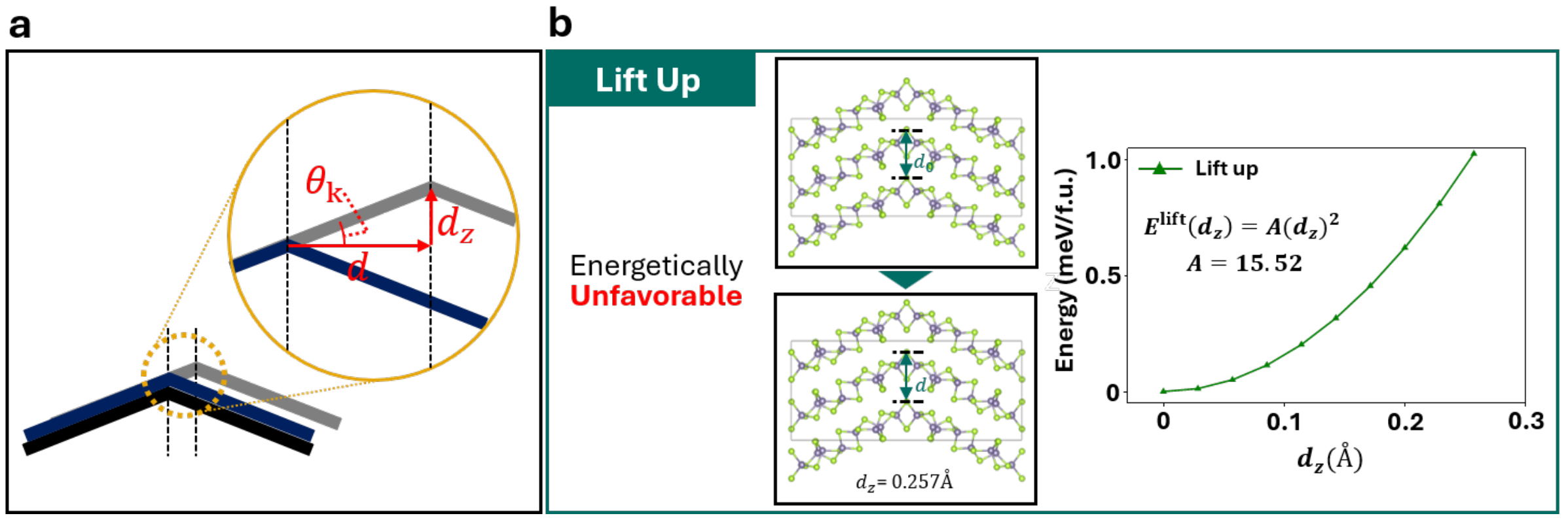

**Figure S15. Geometrical model and DFT evaluation of the interlayer lift-up deformation.** **a,** Schematic illustration of the lift-up deformation induced by interlayer twisting in the inverted-V stacking geometry. The in-plane displacement at the nanowire edge, $d = R\Phi_t$, produces an out-of-plane lift-up displacement $d_z = d \tan\theta_k$, where $R$ is the nanowire radius, $\Phi_t$ is the interlayer twisting angle, and $\theta_k$ denotes the effective inclination angle of the inverted-V structure. **b,** Representative structural models used to evaluate the lift-up deformation and the corresponding Density Functional Theory (DFT)-calculated relative energy as a function of the lift-up displacement $d_z$. The calculated energy was fitted using the quadratic function $\Delta\varepsilon^{\mathrm{lift}}(d_z) = Ad_z^2$, yielding $A = 15.52\ \mathrm{meV/(f.u.\cdot Å^2)}$. The increase in energy with $d_z$ indicates that the lift-up deformation introduces an additional energetic penalty against interlayer twisting.

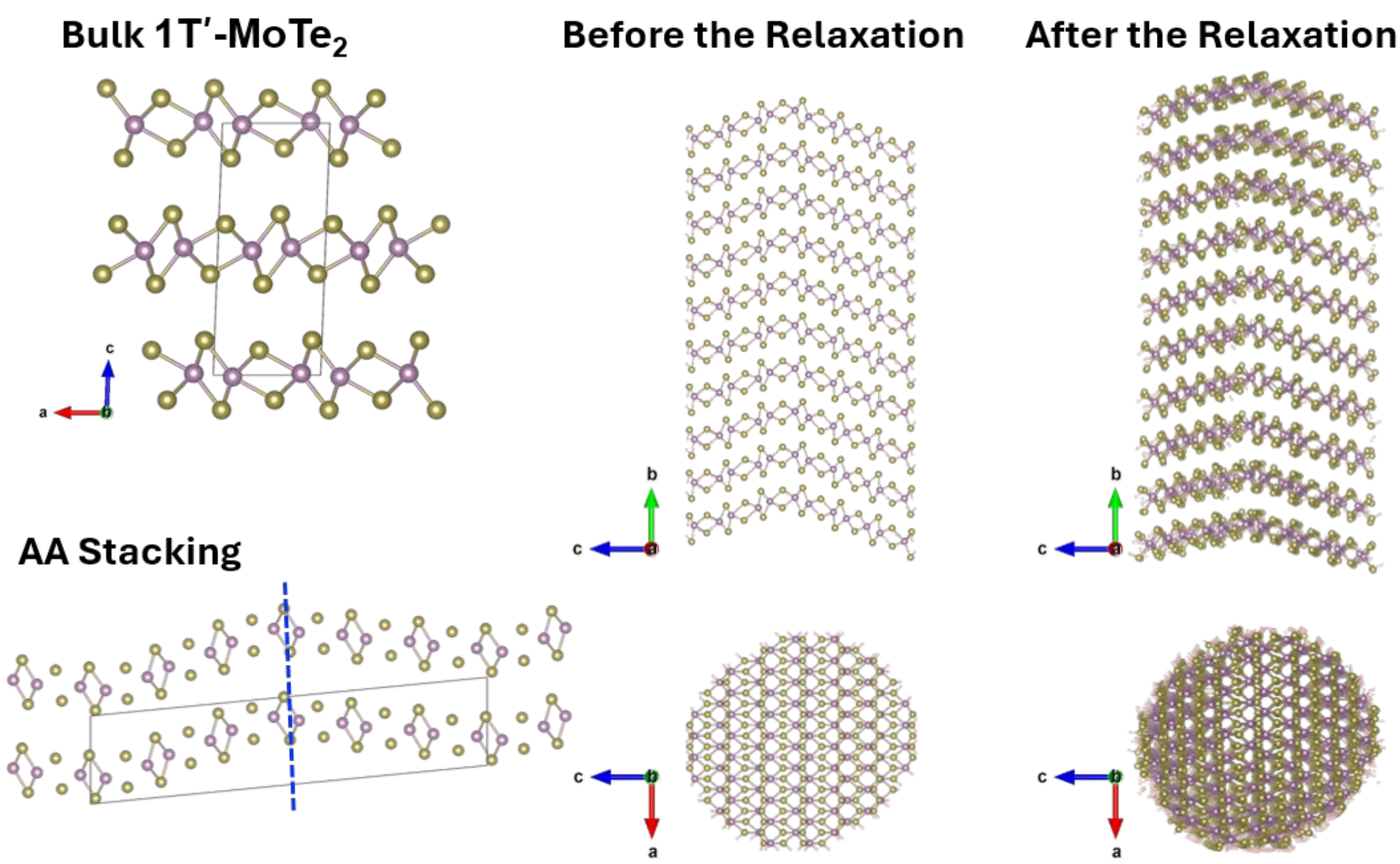

**Figure S16. Structural relaxation simulation of a 1T′-$MoTe_2$ model incorporating a rotational twin boundary.** A cylindrical 1T′-$MoTe_2$ model with a monoclinic structure and rotational twin configuration was constructed to test the applicability of the proposed twisting mechanism. Structural relaxation reproduces continuous layer-by-layer rotation without introducing a screw dislocation, consistent with the mechanism proposed in this study.

**Table S1. Refined unit-cell parameters of the synthesized $GeSe_{2-x}Te_x$ nanowire and comparison with reported β-$GeSe_2$.**

| | This work | β-$GeSe_2$ |
|---|---|---|
| **System** | Triclinic | Monoclinic |
| **Space Group** | $P1$ | $P2_1/c$ |
| **a (Å)** | 7.0 | 7.0 |
| **b (Å)** | 17.0 | 16.8 |
| **c (Å)** | 7.6 | 11.8 |
| **α (°)** | 110.0 | 90.0 |
| **β (°)** | 117.0 | 90.7 |
| **γ (°)** | 90.0 | 90.0 |
| **Volume($Å^3$)** | 739.2 | 1394.1 |
| **# of atoms in the unit cell** | 24 | 48 |

The refined triclinic unit cell of the synthesized nanowire is compared with the previously reported monoclinic β-$GeSe_2$ structure. The structural parameters correspond to the unit cell shown in Fig. 1c.